\documentclass[12pt]{iopart}
\newcommand{\beq}{\begin{equation}} \newcommand{\eeq}{\end{equation}}
\newcommand{\bea}{\begin{eqnarray}} \newcommand{\eea}{\end{eqnarray}}
\newcommand{\bear}{\begin{eqnarray*}} \newcommand{\eear}{\end{eqnarray*}}
\newcommand{\lb}{\label} 
\newcommand{\rf}[1]{(\ref{#1})}   
\usepackage[pdftex]{graphicx}
\usepackage{color}
\usepackage{setstack}
\usepackage{multirow}

\begin{document}

\title{Asymmetric exclusion model with several kinds of impurities.} 

\author{Matheus J. Lazo$^{1}$ and Anderson A. Ferreira$^2$}

\address{$^1$ Instituto de Matem\'atica, Estat\'\i stica e F\'\i sica - FURG, Rio Grande, RS, Brazil.} \ead{matheuslazo@furg.br} 

\address{$^2$ Instituto de F\'{\i}sica e Matem\'atica - Ufpel, Pelotas, RS, Brazil.}

\begin{abstract}

We formulate a new integrable asymmetric exclusion process with $N-1=0,1,2,\dots$ kinds of impurities and with hierarchically ordered dynamics. The model we proposed displays the full spectrum of the simple asymmetric exclusion model plus new levels. The first excited state belongs to these new levels and displays unusual scaling exponents. We conjecture that, while the simple asymmetric exclusion process without impurities belongs to the KPZ universality class with dynamical exponent $\frac{3}{2}$, our model has a scaling exponent $\frac{3}{2}+N-1$. In order to check the conjecture, we solve numerically the Bethe equation with $N=3$ and $N=4$ for the totally asymmetric diffusion and found the dynamical exponents $\frac{7}{2}$ and $\frac{9}{2}$ in these cases. 



\end{abstract}

\maketitle


\section{Introduction}

The simple asymmetric exclusion model (ASEP) is a stochastic model that describes the dynamics of hard-core particles diffusing asymmetrically on the lattice. This model became a paradigm in non-equilibrium statistical physics in the same way that the Ising model in the equilibrium statistic mechanics. Due to its intrinsic nontrivial many-body behavior, the ASEP is used to modeling a wide range of complex systems, like traffic flow \cite{traffic}, biopolymerization \cite{biopolymerization}, interface growth \cite{interface}, etc (see \cite{stochastic} for a review). Remarkably, the ASEP in one-dimension is exactly solvable, what enable us to use the Bethe Ansatz \cite{bethe} to obtain spectral information about its evolution operator \cite{GwaSpohn,Schutz,Kim95,GolinelliMallick}. The relaxation time to the stationary state depends on the system size $L$ and satisfies a scaling relation $T \sim L^z$, where $z=\frac{3}{2}$ is the ASEP dynamical exponent. This dynamic exponent was first obtained by the Bethe Ansatz \cite{GwaSpohn,Kim95,Dhar} and shows that the ASEP belongs to the Kardar-Parizi-Zhang (KPZ) universality class \cite{KPZ}. The scaling property of the model can be understood by mapping the ASEP into the particle height interface model, whose fluctuations in the continuum limit are governed by the KPZ model \cite{KPZ}.

On the other hand, the generalization of the simple exclusion problem by including more than one kind of particles ($N=1,2,...$) has displayed exciting new physics, including spontaneous symmetry breaking and phase separation phenomena \cite{revSchutz}. The introduction of a second class of particle is a useful tool to study the microscopic structure of shocks \cite{BCFNFKS}, and the case with three distinct classes of particles was first considered in \cite{MMR}. However, the critical phenomena and universal dynamics of these one-dimensional driven diffusive systems with several kinds of particles are largely unexplored. Another motivation for studying these models stems from the connection between interacting stochastic particle dynamics and quantum spin systems. This connection follows from the similarity between the master equation describing the time fluctuations of these models and the Schr\"odinger equation in Euclidean time. This relationship enables us to identify a quantum Hamiltonian associated for these stochastic models. The simplest example is the mapping between the ASEP and the exact integrable anisotropic Heisenberg chain, or the so called, XXZ quantum chain \cite{stochastic}. Furthermore, $N$-state quantum Hamiltonians have played an important role in describing strongly correlated electrons in the last decades. Remarkably, in one dimension several models in this category are exactly solvable, as for example, the spin-$1$ Sutherland \cite{Sutherland} and $t$-$J$ \cite{schlo} models, and the spin-$\frac{3}{2}$ Perk-Schultz model \cite{PerkSchultz}, the Essler-Korepin-Schoutens model \cite{Essler}, the Hubbard model \cite{LiebWu} and the two-parameter integrable model introduced in \cite{AlcBar2}. These quantum models can be related to the asymmetric diffusion of two (spin-$1$) and three (spin-$\frac{3}{2}$) kinds of particles \cite{stochastic}, respectively. In its formulation in terms of particles with two and three global conservation laws, these models describe the dynamics of different kinds of particles on the lattice, where the total number of particles of each type is conserved separately. 

In order to ensure integrability, all known models of this class satisfy some particle-particle exchange symmetries \cite{popkov,lazo2}. Recently, we introduced a new class of $3$-state model that is integrable despite it do not have particle-particle exchange symmetry \cite{lazo2}. In \cite{LazoAnd} we extend the model \cite{lazo2} and formulate an one-dimensional asymmetric exclusion process with one kind of impurities (ASEPI). This model describes the dynamics of two types of particles (type $1$ and $2$) on a lattice of $L$ sites, where each lattice site can be occupied by at most one particle. While particles of type $1$ can jump to neighbors sites if they are empty, like in ASEP, particles of type $2$ (called impurities) do not jump to empty sites but exchange positions with neighbor particles of type $1$. We show that this model has a relaxation time longer than the ones for the ASEP, and displays a scaling exponent of $z=\frac{5}{2}$ \cite{LazoAnd} (of order $L^{\frac{3}{2}}\times L=L^{\frac{5}{2}}$ \cite{LazoAnd}). We obtained this result by solving the Bethe Ansatz equation for the half-filling sector and in the totally asymmetric diffusion process \cite{LazoAnd}. 

In the present work we show how this model can be easily generalized to obtain models with relaxation times even larger. We formulate an asymmetrical diffusion model of $N=1,2,3,...$ kinds of particles with impurities (N-ASEPI), where particles of kind $1$ can jump to neighboring sites if they are empty and particles of kind $\alpha=2,3,...,N$ (called impurities) only exchange positions with particles if satisfy a well defined dynamics. Different from the ASEPI \cite{LazoAnd}, our generalized model can have more than one particle on each site (multiple site occupation). Although our model can be solved by the coordinate Bethe ansatz, we are going to formulate a new matrix product ansatz (MPA) \cite{alclazo,popkov} due its simplicity and unifying implementation for arbitrary systems. This new MPA introduced in \cite{alclazo,popkov} can be seen as a matrix product formulation of the coordinate Bethe Ansatz and it is suited to describe all eigenstates of integrable models. We solve this model with periodic boundary condition through the MPA and we analyze the spectral gap for some special cases. Our N-ASEPI model displays the full spectrum of the ASEP \cite{GwaSpohn} plus new levels. The first excited state belongs to these new levels and displays unusual scaling exponents. Although the ASEP belongs to the KPZ universality class, characterized by the dynamical exponent $z = \frac{3}{2}$ \cite{KPZ}, we conjecture that our model displays a scaling exponent $\frac{3}{2}+N-1$, where $N-1$ is the number of kinds of impurities. In order to check our conjecture, we solve numerically the Bethe equation with $N=3$ and $N=4$ for the totally asymmetric diffusion and found that the gap for the first excited state scales with $L^{-\frac{7}{2}}$ and $L^{-\frac{9}{2}}$ in these cases. Furthermore, we also generalize the model \cite{LazoAnd} to include quantum spin chain and solve the Bethe Ansatz equation for symmetric and asymmetric diffusion.

Our paper is organized as follows. In section 2 we generalize the model \cite{LazoAnd} to include quantum spin chain and solve the Bethe Ansatz equation for the symmetric and asymmetric diffusion. The generalization for several kinds of impurities is done in section 3. Finally, our conclusions are presented in Section 4.


\section{The asymmetric exclusion model with one kind of impurities}

        Recently, we propose an exactly solvable asymmetric exclusion process with impurities \cite{LazoAnd} (ASEPI) and found its dynamic exponent $z=\frac{5}{2}$. The exponent $z$ in \cite{LazoAnd} was obtained, from the spectral gap of the model, for the totally asymmetric exclusion process (TASEPI) and at half-filling. It is important to notice that although the ASEP without impurities belongs to the KPZ universality class \cite{KPZ} (dynamic exponent $\frac{3}{2}$), our new model displays an unusual scaling exponent $\frac{5}{2}$. In this section we extend our previous analysis \cite{LazoAnd} and obtain the spectral gap for the symmetric and asymmetric exclusions process. Furthermore, we generalize both the models \cite{lazo2} and \cite{LazoAnd} in order to include quantum spin chains, and we found analytically the spectral gap of the quantum model in the special case where we have free fermions.
        
        The model in \cite{LazoAnd} describes the dynamics of two kinds of particles (type $1$ and $2$) on an one-dimensional lattice of $L$ sites, where each lattice site can be occupied by at most one particle. Furthermore, the total numbers $n_1$, $n_2$ of particles of each type is conserved. In this model if the neighbor sites are empty, particles of type $1$ can jump to the right or to the left with rate $\Gamma_{0\;1}^{1\;0}$ and $\Gamma_{1\;0}^{0\;1}$, respectively. Particles of type $2$ (impurities) do not jump to neighboring sites if they are empty, but can exchange positions with neighbor particles of type $1$ with rates $\Gamma_{2\;1}^{1\;2}$ and $\Gamma_{1\;2}^{2\;1}$ if the the particle $1$ is on the left or on the right, respectively. To describe the occupancy of a given site $i$ ($i=1,2,...,L$), we attach on it a variable $\alpha_i$ taking values $\alpha_i=0,1,2$. If $\alpha_i=0$, the site is vacant. If $\alpha_i=1,2$, we have on the site a particle of kind $1$ or $2$, respectively. The allowed configurations can be denoted by the set $\lbrace\alpha\rbrace=\{ \alpha_1,\alpha_2,...,\alpha_L\}$ of $L$ integers $\alpha_i=0,1,2$. The master equation for the probability distribution at a given time $t$, $P(\lbrace\alpha\rbrace,t)$, can be written in general as
\beq
\label{e0a} 
\frac{\partial P(\lbrace\alpha\rbrace,t)}{\partial t} =  \Gamma(\lbrace\alpha'\rbrace 
\rightarrow \lbrace\alpha\rbrace) P(\lbrace\alpha'\rbrace,t) - \Gamma(\lbrace\alpha\rbrace 
\rightarrow \lbrace\alpha'\rbrace) P(\lbrace\alpha\rbrace,t)
\eeq
where $\Gamma(\lbrace\alpha\rbrace \rightarrow \lbrace\alpha'\rbrace)$ is 
the transition rate where the configuration $\lbrace\alpha\rbrace$ changes to $\lbrace\alpha'\rbrace$.
        
        The master equation \rf{e0a} can be written as a Schr\"odinger equation in Euclidean time (see~\cite{stochastic} for general applications for two-body processes)
\beq 
\label{e0b}
\frac{\partial | P\rangle}{\partial t}=-H | P\rangle,
\eeq
where we represent a configuration $\alpha_i$ on site $i$ by the vector $|\alpha_i\rangle_i$, and we interpret $| P\rangle = P(\lbrace\alpha\rbrace,t)|\alpha_1\rangle\otimes |\alpha_2\rangle\otimes \cdots \otimes|\alpha_L\rangle$ as the associated wave function. In order to generalize our model \cite{LazoAnd} and to include quantum chains solutions, the general Hamiltonian we consider on a ring of perimeter $L$ is given by:
\bea
H&=&\sum_{j=1}^L\left( \Gamma_{0\;1}^{1\;0} E_j^{0,1}E_{j+1}^{1,0}+\Gamma_{1\;0}^{0\;1} E_j^{1,0}E_{j+1}^{0,1}+\sum_{\alpha\neq \beta=1}^2 \Gamma_{\beta\;\alpha}^{\alpha\;\beta} E_j^{\beta,\alpha}E_{j+1}^{\alpha,\beta}\right.\nonumber \\
&& + \left.\sum_{\alpha,\beta=0}^2\Gamma_{\alpha\;\beta}^{\alpha\;\beta} E_j^{\alpha,\alpha}E_{j+1}^{\beta,\beta} \right),\lb{e1}
\eea
with $E^{\alpha,\beta}_{L+1} \equiv E^{\alpha,\beta}_1$ due to the periodic boundary condition, and where $E^{\alpha,\beta}_k$ ($\alpha,\beta=0,1,2$) are the $3\times 3$ Weyl matrix acting on site $k$ with $i,j$ elements $\left(E^{l,m}_k \right)_{i,j}=\delta_{l,i}\delta_{m,j}$, and $\Gamma_{n\;o}^{l\;m}$ are the coupling constants. The last sum in \rf{e1} accounts for the static interactions while the first and second sums are the kinetic terms representing the motion and interchange of particles, respectively. The $U(1)\otimes U(1)$ symmetry supplemented by the periodic boundary condition of \rf{e1} imply that the total number of particles $n_1, n_2=0,1,2...,L$ (with $n_1+n_2 \leq L$) on class $1$ and $2$ as well the momentum $P=\frac{2\pi l}{L}$ ($l=0,1,\ldots,L-1$) are good quantum numbers. Furthermore, the Hamiltonian \rf{e1} also preserves the numbers of vacant sites between the impurities. This conservation plays a fundamental role in the spectrum properties of the model \cite{LazoAnd}. As we shall show, for the stochastic model the Bethe equation do not depends on the number of impurities ($n_2\neq 0$). Consequently, the roots of Bethe equation and the eigenvalues of the Hamiltonian are independent of $n_2\neq 0$ (but the wave function depends on $n_2$). This huge spectrum degeneracy follows directly from the conservation of the numbers of vacant sites between the impurities by the Hamiltonian \rf{e1}. Let us explain with the following example. Suppose we start with a given configuration $0120220$ with one particle ($1$), $3$ impurities ($2$) and $3$ vacant sites ($0$). We can make a surjective map between all possible configuration of these particles to all possible configurations of a new chain with just impurities and vacant sites. For example $0120220 \Longrightarrow 020220$. On this new chain, we are looking only for the effective movement of impurities on the chain. For simplicity, let us consider the totally asymmetric model (TASEP) where $\Gamma_{0\;1}^{1\;0}=1$ and $\Gamma_{1\;0}^{0\;1}=0$. When the particle jumps over the impurities, nothing changes in the effective chain since we also have $0210220 \Longrightarrow 020220$, then $0201220 \Longrightarrow 020220$, then $0202120 \Longrightarrow 020220$, then $0202210 \Longrightarrow 020220$, then $0202201 \Longrightarrow 020220$, and finally a change in the mapped configuration $1202200 \Longrightarrow 202200$. On other words, the impurities move on the mapped chain as they are just one "object" due to the conservation of vacant sites between impurities. Moreover, this "object" only moves when the particle complete a turn over the chain. As a consequence, the time for the particle to complete one revolution is the time scale for the movements of the this "object". For an arbitrary number of particles in a chain of length $L$, the time for the particles to complete one revolution is of order $L^{\frac{3}{2}}$ ($L^2$ in the symmetrical diffusion). As the "object" formed by the impurities need to move of order $L$ times to span all possible configurations, it will takes a time of order $L^{\frac{3}{2}}\times L = L^{\frac{5}{2}}$ ($L^3$ in the symmetrical diffusion) to reach the stationary state.

\subsection{The exact solution of the model}

        We want to formulate a matrix product ansatz for the eigenvectors $|\Psi_{n_1,n_2,P}\rangle$ of the eigenvalue equation
\beq
H|\Psi_{n_1,n_2,P}\rangle=\varepsilon^{n_1,n_2}|\Psi_{n_1,n_2,P}\rangle
\lb{e2}
\eeq
belonging to the eigensector labeled by ($n_1,n_2,P$). These eigenvectors are given by
\beq
|\Psi_{n_1,n_2,P}\rangle=\sum_{\{\alpha \}}\sum_{\{x \}} f(x_1,\alpha_1;\ldots;x_n,\alpha_n)|x_1,\alpha_1;\ldots;x_n,\alpha_n\rangle,
\lb{e3}
\eeq
where the kets $|x_1,\alpha_1;\ldots;s_n,\alpha_n\rangle \equiv \left(|0\rangle\otimes\right)^{x_1-1}|\alpha_1\rangle \otimes \left(|0\rangle\otimes\right)^{x_2-x_1-1} |\alpha_2\rangle \otimes \cdots \otimes \left(|0\rangle\otimes\right)^{x_n-x_{n-1}-1}|\alpha_n\rangle \otimes \left(|0\rangle\otimes\right)^{L-x_n}$ denote the configurations with particles of type $\alpha_i$ ($\alpha_i=1,2$) located at the positions $x_i$ ($x_i=1,\ldots,L$), and the total number of particles is $n=n_1+n_2$. The summation $\{\alpha \}=\{\alpha_1,\ldots,\alpha_n\}$ extends over all the permutations of $n$ integers numbers $\{1,2 \}$ in which $n_1$ terms have value $1$ and $n_2$ terms the value $2$, while the summation $\{x \}=\{x_1,\ldots,x_n \}$ extends, for each permutation $\{\alpha \}$, into the set of the non-decreasing integers satisfying $x_{i+1}\ge x_i+1$.

	The MPA \cite{alclazo} is constructed by making a one-to-one correspondence between the configurations of particles and a product of matrices
\bea
\lb{e4}
f(x_1,\alpha_1;\ldots;x_n,\alpha_n)\Longleftrightarrow && E^{x_1-1}A^{(\alpha_1)}E^{x_2-x_1-1}A^{(\alpha_2)}\cdots \\
&& \;\;\;\;\;\;\;\;\;\; \cdots E^{x_n-x_{n-1}-1}A^{(\alpha_n)}E^{L-x_n}, \nonumber
\eea
where for this map we can choose any operation on the matrix products that give a non-zero scalar. In  the original formulation of the MPA  
with periodic boundary conditions \cite{alclazo} the trace operation was chosen to produce this scalar. The matrices $A^{(\alpha)}$ are associated to the particles of type $\alpha=1,2$, respectively, and the matrix $E$ is associated to the vacant sites. Actually $E$ and $A^{(\alpha)}$ are abstract operators with an associative product. 
A well defined eigenfunction is obtained, apart from a normalization factor, 
if all the amplitudes are related uniquely, due to the algebraic relations (to be fixed) among the matrices $E$ and $A^{(\alpha)}$. Equivalently, the correspondence (\ref{e4}) implies that, in the subset of words (products of matrices) of the algebra containing $n$ matrices $A^{(\alpha)}$ and $L-n$ 
matrices $E$ there exists only a single independent word ("normalization constant"). The relation between any two words is a $c$ number that gives the ratio between the corresponding amplitudes in \rf{e3}.

As the Hamiltonian \rf{e1} commutes with the momentum operator due to the periodic boundary condition, the amplitudes $f(x_1,\alpha_1;\ldots;x_n,\alpha_n)$ should satisfy the following relations:
\beq
\lb{e4b}
f(x_1,\alpha_1;\ldots;x_n,\alpha_n)=e^{-iP}f(x_1+1,\alpha_1;\ldots;x_n+1,\alpha_n),
\eeq
where
\beq
\lb{e4c}
P=\frac{2\pi l}{L}, \; l=0,1,...,L-1.
\eeq
Let us consider initially the simpler cases where $n=1$ and $n=2$.


{\it {\bf  n = 1.}}
We have distinct equations depending on the type $\alpha=1,2$ of the particle. The eigenvalue equation \rf{e2} give us
\bea
&&\varepsilon^{(1)} E^{x-1}A^{(1)}E^{L-x}=\Gamma_{0\;1}^{1\;0} E^{x-2}A^{(1)}E^{L-x+1}+\Gamma_{1\;0}^{0\;1} E^{x}A^{(1)}E^{L-x-1} \nonumber \\
&&+ \left(\Gamma_{0\;1}^{0\;1}+\Gamma_{1\;0}^{1\;0} \right) E^{x-1}A^{(1)}E^{L-x}, 
\lb{e5}
\eea
if the particle is of type $1$ and
\beq
\varepsilon^{(2)} E^{x-1}A^{(2)}E^{L-x}=\left(\Gamma_{0\;2}^{0\;2}+\Gamma_{2\;0}^{2\;0} \right) E^{x-1}A^{(2)}E^{L-x},
\lb{e6}
\eeq
if the particle is of type $2$. In these last two equations $\varepsilon^{(1)}\equiv \varepsilon^{1,0}$ and $\varepsilon^{(2)}\equiv \varepsilon^{0,1}$ are the eigenvalues, and we choose $\Gamma_{0\;0}^{0\;0}=0$ without loss of generality. A convenient solution is obtained by introducing the spectral parameter dependent matrices
\beq
A^{(\alpha)}=EA_k^{(\alpha)} \;\;\; (\alpha=1,2),
\lb{e7}
\eeq
with complex $k$ parameter, that satisfy the commutation relation with the matrix $E$
\beq
EA_k^{(\alpha)}=e^{ik}A_k^{(\alpha)}E \;\;\; (\alpha=1,2).
\lb{e8}
\eeq
Inserting \rf{e7} and \rf{e8} into \rf{e5} and \rf{e6} we obtain
\bea
&&\varepsilon^{(1)}(k)=\Gamma_{0\;1}^{1\;0}e^{-ik}+\Gamma_{1\;0}^{0\;1}e^{ik}+\Gamma_{0\;1}^{0\;1}+\Gamma_{1\;0}^{1\;0},\nonumber \\
&&\varepsilon^{(2)}(k)=\Gamma_{0\;2}^{0\;2}+\Gamma_{2\;0}^{2\;0}.
\lb{e9}
\eea
The up to now free spectral parameter $k$ is fixed by imposing the boundary condition. This will be done only for general $n$.


{\it {\bf  n = 2.}}
For two particles of types $\alpha_1$ and $\alpha_2$ ($\alpha_1,\alpha_2=1,2$) on the lattice we have two kinds of relations coming from the eigenvalue equation. The configurations where the particles are at positions ($x_1$, $x_2$) with $x_2>x_1+1$ give us the generalization of \rf{e5}
\bea
&&\varepsilon^{n_1,n_2} E^{x_1-1}A^{(\alpha_1)}E^{x_2-x_1-1}A^{(\alpha_2)}E^{L-x_2}= \nonumber \\
&&\Gamma_{0\;\alpha_1}^{\alpha_1\;0} E^{x_1-2}A^{(\alpha_1)}E^{x_2-x_1}A^{(\alpha_2)}E^{L-x_2}+\Gamma_{\alpha_1\;0}^{0\;\alpha_1} E^{x_1}A^{(\alpha_1)}E^{x_2-x_1-2}A^{(\alpha_2)}E^{L-x_2} \nonumber \\
&&+\Gamma_{0\;\alpha_2}^{\alpha_2\;0} E^{x_1-1}A^{(\alpha_1)}E^{x_2-x_1-2}A^{(\alpha_2)}E^{L-x_2+1}\nonumber \\
&& +\Gamma_{\alpha_2\;0}^{0\;\alpha_2} E^{x_1-1}A^{(\alpha_1)}E^{x_2-x_1}A^{(\alpha_2)}E^{L-x_2-1} \nonumber \\
&&+ \left(\Gamma_{0\;\alpha_1}^{0\;\alpha_1}+\Gamma_{\alpha_1\;0}^{\alpha_1\;0} +\Gamma_{0\;\alpha_2}^{0\;\alpha_2}+\Gamma_{\alpha_2\;0}^{\alpha_2\;0}\right) E^{x_1-1}A^{(\alpha_1)}E^{x_2-x_1-1}A^{(\alpha_2)}E^{L-x_2},
\lb{e10}
\eea
and the configurations where the particles are at the colliding positions ($x_1=x$, $x_2=x+1$) give us
\bea
&&\varepsilon^{n_1,n_2} E^{x-1}A^{(\alpha_1)}A^{(\alpha_2)}E^{L-x-1}= \Gamma_{0\;\alpha_1}^{\alpha_1\;0}E^{x-2}A^{(\alpha_1)}EA^{(\alpha_2)}E^{L-x-1} \nonumber \\
&&+\Gamma_{\alpha_2\;0}^{0\;\alpha_2}E^{x-1}A^{(\alpha_1)}EA^{(\alpha_2)}E^{L-x-2} +\Gamma_{\alpha_1\;\alpha_2}^{\alpha_2\;\alpha_1}E^{x-1}A^{(\alpha_2)}A^{(\alpha_1)}E^{L-x-1} \nonumber \\
&&+ \left(\Gamma_{0\;\alpha_1}^{0\;\alpha_1}+\Gamma_{\alpha_2\;0}^{\alpha_2\;0} +\Gamma_{\alpha_1\;\alpha_2}^{\alpha_1\;\alpha_2}\right) E^{x-1}A^{(\alpha_1)}A^{(\alpha_2)}E^{L-x-1},
\lb{e11}
\eea
where we introduced $\Gamma_{0\;2}^{2\;0}=\Gamma_{2\;0}^{0\;2}=0$, and $\Gamma_{\alpha_1\;\alpha_2}^{\alpha_2\;\alpha_1}=0$ if $\alpha_1=\alpha_2$. The Hamiltonian \rf{e1} do not have a standard solution as in \cite{alclazo} where each of the matrices $A^{(\alpha)}$ ($\alpha=1,2$) are composed by two spectral parameter matrices, with the same value of the spectral parameters $k_1,k_2$ (case a in \cite{alclazo}). In order to obtain a solution for \rf{e10}-\rf{e11} we now need to consider the $A^{(\alpha)}$ as composed by $n_{\alpha}$ spectral parameter dependent matrices $A_{k_j^{(\alpha)}}^{(\alpha)}$ belonging to two distinct sets of spectral parameters \cite{lazo2,LazoAnd}, i. e.,
\beq
A^{(\alpha)}=\sum_{j=1}^{n_{\alpha}} EA_{k_j^{(\alpha)}}^{(\alpha)} \;\;\; \mbox{with} \;\;\; EA_{k_j^{(\alpha)}}^{(\alpha)}=e^{ik_j^{(\alpha)}}A_{k_j^{(\alpha)}}^{(\alpha)}E, \;\;\; \left(A_{k_j^{(\alpha)}}^{(\alpha)} \right)^2=0,
\lb{e12}
\eeq
for $\alpha=1,2$ and $n_1+n_2=n$. These last relations when inserted in \rf{e10} give us the energy in terms of the spectral parameters $k_j^{(\alpha)}$ ($\alpha=1,2$)
\beq
\varepsilon^{n_1,n_2}=\sum_{j=1}^{n_1}\varepsilon^{(1)}(k_j^{(1)})+\sum_{j=1}^{n_2}\varepsilon^{(2)}(k_j^{(2)}),
\lb{e13}
\eeq
where $\varepsilon^{(\alpha)}(k)$ is given by \rf{e9}. 

Let us consider now \rf{e11} in the case where the particles are of the same type. For two particles, when $\alpha_1=\alpha_2=1$, the equations \rf{e12}, \rf{e13} and \rf{e11} implies that the matrices $\{ A_{k_j^{(1)}}^{(1)}\}$ should obey the Zamolodchikov algebra \cite{Zamolodchikov}
\beq
A_{k_j^{(1)}}^{(1)}A_{k_l^{(1)}}^{(1)}=S_{1\;1}^{1\;1}(k_j^{(1)},k_l^{(1)})A_{k_l^{(1)}}^{(1)}A_{k_j^{(1)}}^{(1)} \;\;\; (j \ne l), \;\;\; \left(A_{k_j^{(1)}}^{(1)} \right)^2=0,
\lb{e14}
\eeq
where $j,l =1,...,n_1$, and the algebraic constants $S_{1\;1}^{1\;1}(k_j^{(1)},k_l^{(1)})$ are given by:
\beq
S_{1\;1}^{1\;1}(k_j^{(1)},k_l^{(1)})=-\frac{\Gamma_{0\;1}^{1\;0}+\Gamma_{1\;0}^{0\;1}e^{i(k_j^{(1)}+k_l^{(1)})}-\left(\Gamma_{1\;1}^{1\;1}-\Gamma_{1\;0}^{1\;0}-\Gamma_{0\;1}^{0\;1} \right)e^{ik_j^{(1)}}}{\Gamma_{0\;1}^{1\;0}+\Gamma_{1\;0}^{0\;1}e^{i(k_j^{(1)}+k_l^{(1)})}-\left(\Gamma_{1\;1}^{1\;1}-\Gamma_{1\;0}^{1\;0}-\Gamma_{0\;1}^{0\;1} \right)e^{ik_l^{(1)}}}.
\lb{e15}
\eeq
For two impurities ($\alpha_1=\alpha_2=2$) at "colliding" positions, the eigenvalue equation does not fix a commutation relation among the matrices $A_{k_j^{(2)}}^{(2)}$ since \rf{e11} is automatically satisfied in this case. On the other hand, the sum
\beq
\sum_{j,l=1}^2 A_{k_j^{(2)}}^{(2)}E^dA_{k_l^{(2)}}^{(2)}\neq 0,
\lb{e16}
\eeq
where the number of vacant sites between the impurities $d=y-x$ is a conserved charge of the Hamiltonian \rf{e1}, should be different from zero or the MPA will produces an eigenfunction with null norm. Moreover, the algebraic expression in \rf{e12} assures that any matrix product defining our ansatz \rf{e4} can be expressed in terms of two single matrix products $A_{k_1^{(2)}}^{(2)}A_{k_2^{(2)}}^{(2)}E^L$ and $A_{k_2^{(2)}}^{(2)}A_{k_1^{(2)}}^{(2)}E^L$. Using \rf{e12} we have, from the periodic boundary condition,
\beq
A_{k_j^{(2)}}^{(2)}A_{k_l^{(2)}}^{(2)}E^L=e^{-ik_j^{(2)}L}e^{-ik_l^{(2)}L}A_{k_j^{(2)}}^{(2)}A_{k_l^{(2)}}^{(2)}E^L,
\lb{e19}
\eeq
To satisfy this equation we should have $k_2^{(2)}=-k_1^{(2)} + 2\pi j/L$ ($j=0,1,...,L-1$). Consequently, the most general commutation relation $A_{k_1^{(2)}}^{(2)}A_{k_2^{(2)}}^{(2)}=S_{2\;2}^{2\;2}(k_j^{(2)},k_l^{(2)})A_{k_2^{(2)}}^{(2)}A_{k_1^{(2)}}^{(2)}$ among the matrices $A_{k_1^{(2)}}^{(2)}$ and $A_{k_2^{(2)}}^{(2)}$ can be reduced to $A_{k_1^{(2)}}^{(2)}A_{k_2^{(2)}}^{(2)}=A_{k_2^{(2)}}^{(2)}A_{k_1^{(2)}}^{(2)}$ ($S_{2\;2}^{2\;2}(k_j^{(2)},k_l^{(2)})=1$) by an appropriate change of variable in the spectral parameter $k_1^{(2)}$. By choosing $S_{2\;2}^{2\;2}(k_j^{(2)},k_l^{(2)})=1$ and imposing that the sum \rf{e16} is not zero, we obtain ($j,l,v=1,...,n_2$)
\beq
k_l^{(2)}\neq k_j^{(2)} + \frac{\pi (2m+1)}{d_v} \;\; (m=0,1,...),
\lb{e23}
\eeq
where $\{d_v\}$ is the set of all numbers of vacant sites between the impurities.

	Let us consider now the case where the particles are of distinct kinds. From \rf{e9}, \rf{e12} and \rf{e13}, equation \rf{e11} give us two independent relations:
\bea
&&\left[ \Gamma_{0\;\alpha_2}^{\alpha_2\;0}+\Gamma_{\alpha_1\;0}^{0\;\alpha_1}e^{i(k^{(1)}+k^{(2)})} -\left(\Gamma_{\alpha_1\;\alpha_2}^{\alpha_1\;\alpha_2}-\Gamma_{\alpha_1\;0}^{\alpha_1\;0}-\Gamma_{0\;\alpha_2}^{0\;\alpha_2}\right)e^{ik^{(\alpha_2)}} \right]A_{k^{(\alpha_1)}}^{(\alpha_1)}A_{k^{(\alpha_2)}}^{(\alpha_2)} \nonumber \\
&&-\Gamma_{\alpha_1\;\alpha_2}^{\alpha_2\;\alpha_1}e^{ik^{(\alpha_2)}}A_{k^{(\alpha_2)}}^{(\alpha_2)}A_{k^{(\alpha_1)}}^{(\alpha_1)}=0 \;\;\; (\alpha_1 \ne \alpha_2=1,2).
\lb{e27}
\eea
This two relations need to be identically satisfied, and since at this level we want to keep $k^{(1)}$ and $k^{(2)}$ as free complex parameters, \rf{e27} imply special choices of the coupling constants $\Gamma_{k\;l}^{m\;n}$ \cite{lazo2,LazoAnd}:
\beq
\lb{e28}
\Gamma_{2\;0}^{0\;2}=\Gamma_{0\;2}^{2\;0}=0, \;\;\;\;\; \Gamma_{1\;2}^{2\;1}\Gamma_{2\;1}^{1\;2}=\Gamma_{1\;0}^{0\;1}\Gamma_{0\;1}^{1\;0},\;\;\;\;\; t_{12}=t_{21}=t_{22}=0,
\eeq
where $t_{\alpha_1 \alpha_2}=\Gamma_{\alpha_1\;\alpha_2}^{\alpha_1\;\alpha_2}-\Gamma_{\alpha_1\;0}^{\alpha_1\;0}-\Gamma_{0\;\alpha_2}^{0\;\alpha_2}$ ($\alpha_1,\alpha_2=1,2$). We also obtain the structural constants:
\beq
\lb{e29}
S^{2 \; 1}_{2 \; 1}(k^{(2)},k^{(1)})=\frac{1}{S^{1 \; 2}_{1 \; 2}(k^{(1)},k^{(2)})}=\frac{\Gamma_{1\;2}^{2\;1}}{\Gamma_{1\;0}^{0\;1}}e^{ik^{(2)}}.
\eeq
The integrability conditions \rf{e28} generalizes the result obtained for the stochastic process with one kind of impurities \cite{LazoAnd} to include quantum chains \cite{lazo2}. Let us consider now the case of general $n$.


{\it {\bf  General n.}} 
We now consider the case of arbitrary numbers $n_1$, $n_2$ of particles of type $1$ and $2$. The eigenvalue equation gives us generalizations of \rf{e10} and \rf{e11}. To solve these equations we identify the matrices $A^{(\alpha)}$ as composed by $n_{\alpha}$ spectral dependent matrices \rf{e12}. The configurations where $x_{i+1}>x_i+1$ give us the energy \rf{e13}. The amplitudes in \rf{e11} where a pair of particles of types $\alpha_1$ and $\alpha_2$ are located at the closest positions give us the algebraic relations
\beq
\lb{e30}
A_{k_j^{(\alpha_1)}}^{(\alpha_1)}A_{k_l^{(\alpha_2)}}^{(\alpha_2)}=S_{\alpha_1\;\alpha_2}^{\alpha_1\;\alpha_2}(k_j^{(\alpha_1)},k_l^{(\alpha_2)})A_{k_l^{(\alpha_2)}}^{(\alpha_2)}A_{k_j^{(\alpha_1)}}^{(\alpha_1)},
\eeq
where the algebraic structure constants are the diagonal $S$-matrix defined by \rf{e15}, \rf{e29}, $S_{2\;2}^{2\;2}(k_j^{(2)},k_l^{(2)})=1$, with coupling constants \rf{e28}.

        In order to complete our solutions we should fix the spectral parameters $k_1^{(1)},\ldots, k_{n_1}^{(1)}$ and $k_1^{(2)},\ldots, k_{n_2}^{(2)}$. The algebraic expression in \rf{e12} assures that any matrix product defining our ansatz \rf{e4} can be expressed in terms of the matrix product $A_{k_1^{(1)}}^{(1)}\cdots  A_{k_{n_1}^{(1)}}^{(1)}A_{k_1^{(2)}}^{(2)}\cdots A_{k_{n_2}^{(2)}}^{(2)} E^L$. From the periodic boundary condition we obtain:
\bea
&&e^{ik_j^{(1)}L}=-e^{-i\sum_{l=1}^{n_2}k_l^{(2)}}\left(\frac{\Gamma_{1\;2}^{2\;1}}{\Gamma_{1\;0}^{0\;1}}\right)^{-n_2}\prod_{l=1}^{n_1}S_{1\;1}^{1\;1}(k_j^{(1)},k_l^{(1)})\;\;\;\; (j=1,...,n_1),\nonumber\\
&&e^{ik_j^{(2)}(L-n_1)}=\left(\frac{\Gamma_{1\;2}^{2\;1}}{\Gamma_{1\;0}^{0\;1}}\right)^{n_1}\;\;\;\;(j=1,...,n_2),
\lb{e36}
\eea 
where the spectral parameters $\{k_j^{(2)}\}$ should satisfy the restrictions \rf{e23}. We can write the Bethe equation \rf{e36} in a more convenient way. From the second expression on \rf{e36} we have
\beq
e^{i\sum_{j=1}^{n_2}k_j^{(2)}(L-n_1)}=\left(\frac{\Gamma_{1\;2}^{2\;1}}{\Gamma_{1\;0}^{0\;1}}\right)^{n_1n_2} \Rightarrow e^{i\sum_{j=1}^{n_2}k_j^{(2)}}=\left(\frac{\Gamma_{1\;2}^{2\;1}}{\Gamma_{1\;0}^{0\;1}}\right)^{\frac{n_1n_2}{L-n_1}}e^{i\frac{2\pi}{L-n1}m},
\lb{e38}
\eeq
with $m=0,1,...,L-n_1-1$. By inserting \rf{e38} and using \rf{e15} in the first equation in \rf{e36} we obtain
\beq
e^{ik_j^{(1)}L}=(-)^{n_1-1}\phi_{n_1,n_2}(m)\prod_{l=1}^{n_1}\frac{\Gamma_{0\;1}^{1\;0}+\Gamma^{0\;1}_{1\;0}e^{i(k_j^{(1)}+k_l^{(1)})}-2\Delta e^{ik_j^{(1)}}}{\Gamma_{0\;1}^{1\;0}+\Gamma^{0\;1}_{1\;0}e^{i(k_j^{(1)}+k_l^{(1)})}-2\Delta e^{ik_l^{(1)}}},
\lb{e39}
\eeq
where $j =1,...,n_1$, $2\Delta=\Gamma_{0\;0}^{0\;0}+\Gamma_{1\;1}^{1\;1}-\Gamma_{1\;0}^{1\;0}-\Gamma_{0\;1}^{0\;1}$, and the phase factor $\phi_{n_1,n_2}(m)$ is defined by:
\beq
\phi_{n_1,n_2}(m)=\left(\frac{\Gamma_{1\;2}^{2\;1}}{\Gamma_{1\;0}^{0\;1}}\right)^{\frac{-n_1(n_2)^2}{L-n_1}}e^{-i\frac{2\pi}{L-n_1}m} \;\;\; (m=0,1,...,L-n_1-1).
\lb{e40} 
\eeq
The Bethe equation \rf{e39} generalizes our results \cite{lazo2,LazoAnd}. Furthermore, it is important to notice that \rf{e40} differs from the ones related to asymmetric XXZ chain \cite{GwaSpohn} by the phase factor $\phi_{n_1,n_2}(m)$ \rf{e40}. As we shall see, this phase factor will play a fundamental role in the spectral properties of the model.

Finally, the eigenstate momentum is given by inserting the ansatz \rf{e12} into the relation \rf{e4b}:
\beq
\lb{e35b}
P=\sum_{j=1}^{n_1} k_j^{(1)} + \sum_{j=1}^{n_2} k_j^{(2)}=\frac{2\pi l}{L} \;\;\; (l=0,1,\ldots,L-1).
\eeq
The Bethe equation \rf{e40} plus the momentum equation \rf{e35b} completely fix the spectral parameters $\{k_j^{(1)}\}$ and $\{k_j^{(2)}\}$ and the eigenvalues \rf{e13}. Let us consider some special cases:

\subsection{Stochastic Model} For stochastic models we should have $\Gamma_{\beta\;\alpha}^{\alpha\;\beta}=-\Gamma_{\alpha\;\beta}^{\alpha\;\beta}$ ($\alpha,\beta=0,1,2$). Let set, without loss of generality, $\Gamma_{0\;1}^{1\;0}+\Gamma^{0\;1}_{1\;0}=1$, $\Gamma_{1\;2}^{2\;1}=\Gamma_{1\;0}^{0\;1}$ and $\Delta=\frac{1}{2}$. In this case our model describes an asymmetric exclusion process with impurities \cite{LazoAnd}. The Bethe equation \rf{e39} with the phase factor \rf{e40} reduces now to
\beq
\lb{e41}
e^{ik_j^{(1)}L}=(-)^{n_1-1}e^{-i\frac{2\pi}{L-n_1}m}\prod_{l=1}^{n_1}\frac{\Gamma_{0\;1}^{1\;0}+\Gamma^{0\;1}_{1\;0}e^{i(k_j^{(1)}+k_l^{(1)})}-e^{ik_j^{(1)}}}{\Gamma_{0\;1}^{1\;0}+\Gamma^{0\;1}_{1\;0}e^{i(k_j^{(1)}+k_l^{(1)})}- e^{ik_l^{(1)}}},
\eeq
where $j =1,...,n_1$, and $m=0,1,...,L-n_1-1$. 

In our previous work \cite{LazoAnd} we consider the TASEPI (when $\Gamma_{0\;1}^{1\;0}=1$ and $\Gamma^{0\;1}_{1\;0}=0$, or $\Gamma_{0\;1}^{1\;0}=0$ and $\Gamma^{0\;1}_{1\;0}=1$). In this case we solve \rf{e41} numerically up to $L=1024$ in the half-filling sector $n_1=L/2$ and we obtain the scaling exponent $z=\frac{5}{2}$ for the TASEPI. Now, we generalize our previous results by solving the Bethe equation \rf{e36} for the asymmetric ASEPI and symmetric SEPI exclusion process. We also checked the eigenvalues obtained from exact diagonalization of the Hamiltonian with the Bethe Ansatz solution for a small chain with $L=6$, $n_1=2$, $n_2=1$ and $\Gamma_{0\;1}^{1\;0}=0.75$ (see Appendix A). The eigenvalue with the largest real part is $\varepsilon^{n_1,n_2}=0$ corresponding to the stationary state (it is provided by choosing $m=0$ and $P=0$ in the Bethe equation \rf{e41} giving us the $n_1$ fugacities $e^{ik_j^{(1)}}=1$). Others eigenvalues contribute to the relaxation behavior to the stationary state. In special, the eigenvalue with the second largest real part determines the relaxation time and the dynamical exponent $z$. This eigenvalue is obtained from the Bethe equation \rf{e41} by choosing $m=1$ and $P=\frac{2\pi}{L}$ (see Appendix A for a detailed discussion for a small chain). In Table \rf{table2} we show the dynamical exponent $z$ versus $\Gamma_{0\;1}^{1\;0}$ obtained from the numerical solution of the Bethe equation \rf{e36} for several values of $L$ in the half-filling sector $n_1=L/2$. The errors displayed are computed from the linear regression for the logarithm of the real part of the energy gap versus the logarithm of $L$ (see Figure \rf{fig0} for the cases where $\Gamma_{0\;1}^{1\;0}=0.7$ and $\Gamma_{0\;1}^{1\;0}=0.5$).
\begin{figure}[ht]
\includegraphics[width=\textwidth]{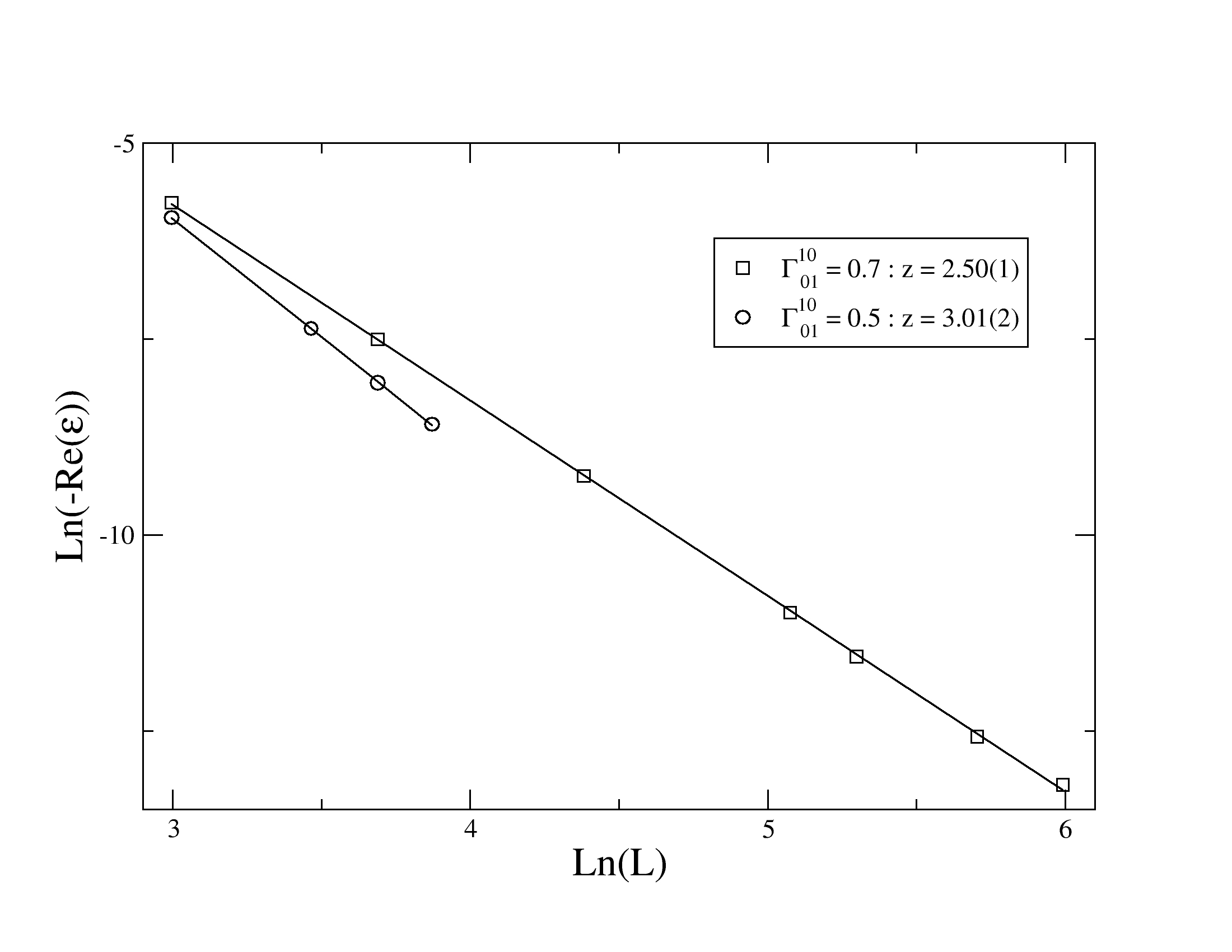}
\caption{In this figure we display the logarithm of the real part of the energy gap versus the logarithm of $L$ for $\Gamma_{0\;1}^{1\;0}=0.7$ and $\Gamma_{0\;1}^{1\;0}=0.5$. The dynamical exponents $z$ are $2.50(2)$ and $3.01(2)$, respectively.}\label{fig0}
\end{figure} 
\begin{table}[ht]
\caption{Dynamical exponent $z$ versus $\Gamma_{0\;1}^{1\;0}$ in the half-filling sector $n_1=L/2$}\label{table2}
\vspace{3mm}
\centering
\begin{tabular}{|c|c|c|c|c|c|c|}
\hline
   & $\Gamma_{0\;1}^{1\;0}=0.9$ & $\Gamma_{0\;1}^{1\;0}=0.8$& $\Gamma_{0\;1}^{1\;0}=0.7$ & $\Gamma_{0\;1}^{1\;0}=0.6$ & $\Gamma_{0\;1}^{1\;0}=0.5$ \\ 
\hline
$z$ & 2.52(2) & 2.51(2) & 2.50(1) & 2.52(2) & 3.01(2)\\ 
\hline
\end{tabular}
\end{table}
In the ASEPI, we consider the cases where $\Gamma_{0\;1}^{1\;0}=0.9,0.8,0.7,0.6$ with $L=20,40,80,160,200,300,400$, and we found that the energy gap has a leading behavior of KPZ $L^{-\frac{3}{2}}$ and a sub-leading term $L^{-\frac{5}{2}}$ related to the super-diffusion of particles $1$ and sub-diffusion of particles $2$, respectively. For $m=0$ (and also for $n_2=0$) the sub-leading term vanishes and we recover the spectrum of the ASEP without impurities \cite{GwaSpohn}. For $m=1$ the spectral gap scales with $L^{-\frac{5}{2}}$ instead of $L^{-\frac{3}{2}}$ due to the sub-leading term. On the other hand, we obtain for the SEPI ($\Gamma_{0\;1}^{1\;0}=\Gamma^{0\;1}_{1\;0}=0.5$ with $L=20,32,40,52$), a energy gap with a leading behavior of $L^{-2}$ and a sub-leading term $L^{-3}$ related to the the normal diffusion of particles $1$ and sub-diffusion of particles $2$, respectively.

Finally, as in \cite{LazoAnd}, it is important to notice that the dynamics of particles $1$ is not affected by the impurities. This explain why the model displays the full spectrum of the ASEP \cite{GwaSpohn}. The dynamics of the impurities is totally dependent on the particles, since the impurities only move when particles change position with them. Consequently, the time to vanish the fluctuations on the densities of particles acts as a time scale for the diffusion of the impurities, resulting in a relaxation time greater than the one for the standard ASEP and SEP (reflected in the $L^{-\frac{3}{2}}\times L^{-1} = L^{-\frac{5}{2}}$ gap for the ASEPI and $L^{-2}\times L^{-1} = L^{-3}$ gap for the SEPI).

{\bf Eigenstates:} The stationary state is the eigenstate associated to the eigenvalue $\varepsilon^{n_1,n_2}=0$ (it is in the sector with $m=0$ and $P=0$). In this case, the Bethe equation \rf{e41} has the unique solution $e^{k_{j}^{(\alpha)}}=1$ for all $\alpha=1,2$ and $j=1,...,n_{\alpha}$. As a consequence, the $S$-matrix reduces to the identity $S_{\alpha_1\;\alpha_2}^{\alpha_1\;\alpha_2}(k_j^{(\alpha_1)},k_l^{(\alpha_2)})=1$, and all amplitudes in the eigenfunction \rf{e3} becomes equals to a normalization constant $f(x_1,\alpha_1;\ldots;x_n,\alpha_n)=f_0$, since we have from \rf{e4} and \rf{e12}:
\bea
\lb{e41b}
E^{x_1}A^{(\alpha_1)}_{k^{\alpha_1}_1}E^{x_2-x_1}A^{(\alpha_2)}_{k^{\alpha_2}_2}\cdots E^{x_n-x_{n-1}}A^{(\alpha_n)}_{k^{\alpha_n}_n}E^{L-x_n}\\
\;\;\;\;\;\;\;\;\;\;\;\;\;\;\;\;\;\;\;\;\;\;\;\;\;\;\;\;\;\;\;\;\;\;\;\;\;\;\;\;=A_{k_1^{(1)}}^{(1)}\cdots  A_{k_{n_1}^{(1)}}^{(1)}A_{k_1^{(2)}}^{(2)}\cdots A_{k_{n_2}^{(2)}}^{(2)} E^L. \nonumber
\eea
The eigenfunction $|\Psi_{0}\rangle$ corresponding to the stationary state is given by a simple combination of all possible configurations of particles, where each particle configuration has the same weight given by the normalization constant $f_0$. We have from \rf{e3}:
\beq
|\Psi_{0}\rangle=f_0\sum_{\{\alpha \}}\sum_{\{x \}} |x_1,\alpha_1;\ldots;x_n,\alpha_n\rangle.
\lb{e41c}
\eeq
Consequently, at the stationary state, all configurations that satisfy the hard-core constraints imposed by the definition of the model occur with equal probabilities given by $f_0$. Furthermore, in this case each site is occupied by a particle $\alpha$ with probability $\rho_{\alpha}=n_{\alpha}/L$.

Finally, the MPA \rf{e4} enable us to write all eigenstates of \rf{e1} in a matrix product form. For a given solution $k_j^{(1)}$ and $k_j^{(2)}$, the matrices $E$ and $A_{k_j^{(\alpha)}}^{(\alpha)}$ have the following finite-dimensional representation:
\bea
\lb{e42}
E&=&\bigotimes_{l=1}^{n_1} \left( \begin{array}{cc}
				  1 & 0 \\
                                  0 & e^{ik_l^{(1)}}
			          \end{array}
			  	 \right)
				  \bigotimes_{l=1}^{n_2} \left( \begin{array}{cc}
				  1 & 0 \\
                                  0 & e^{ik^{(2)}_l}
			          \end{array}
			          \right), \\
A^{(2)}_{k^{(2)}_j}&=&\bigotimes_{l=1}^{n_1} \left( \begin{array}{cc}
				  1 & 0 \\
                                  0 & \frac{\Gamma_{1\;2}^{2\;1}}{\Gamma_{1\;0}^{0\;1}}e^{ik^{(2)}_l}
			          \end{array}
			  \right)
		    \bigotimes_{l=1}^{j-1} I_2\bigotimes
				 \left( \begin{array}{cc}
				  0 & 0 \\
                                  1 & 0
			          \end{array}
			  \right)
			\bigotimes_{l=j+1}^{n_2} I_2,\nonumber \\
A^{(1)}_{k_j^{(1)}}&=&\bigotimes_{l=1}^{j-1} \left( \begin{array}{cc}
				  S^{1 \; 1}_{1 \; 1}(k_j^{(1)},k_l^{(1)}) & 0 \\
                                  0 & 1
			          \end{array}
			  \right)\bigotimes\left( \begin{array}{cc}
				  0 & 0 \\
                                  1 & 0
			          \end{array}
			  \right)
		    \bigotimes_{j=j+1}^{n_1+n_2} I_2,
\nonumber
\eea
where $n=n_1+n_2$, $I_2$ is the $2\times 2$ identity matrix, and the dimension of the representation is $2^n$. The matrix product form of the eigenstates are given by inserting the matrices \rf{e12} defining the MPA \rf{e4} with the spectral parameters \rf{e39} into equation \rf{e3}. As showed in \cite{mallick}, our MPA generalizes the steady-state Matrix Product introduced by Derrida et al \cite{Derrida}. For the stochastic model the stationary state is obtained by choosing $k_j^{(1)}=0$ and $k_j^{(2)}=0$ in \rf{e42}. However, a relation between our matrix product form for the steady-state and the standard Matrix Product form for open boundary condition is not trivial \cite{mallick,Derrida}.

\subsection{Quantum model} We consider here only the simplest case of free fermions, when $\Delta=0$ and $\Gamma^{2\;1}_{1\;2}=\Gamma_{0\;1}^{1\;0}=\Gamma_{1\;0}^{0\;1}= 1$, in this particular case, the Bethe equation \rf{e39} reduces to roots of unit. The simplicity of Bethe equation enable us to calculate analytically the eigenvalues by following \cite{GwaSpohn}. As in the stochastic model, the ground state is obtained by choosing $m=0$ and the first excited state is given by $m=1$. Due to the $L^{-1}$ term in the phase factor, the energy gap
\beq
\lb{e39b}
\Delta \varepsilon=\frac{4\alpha\pi}{(1-\rho_1)^2}\sin{\left(\frac{\alpha\pi\rho_1}{2}\right)}\frac{1}{L^3}+{\mbox{O}}(L^{-4}),
\eeq
with $\alpha=1(2)$ for $n_1$ even (odd), scales with $L^{-3}$ instead of $L^{-1}$ for the case of only one kind of particle. The boundary condition plays a fundamental role in the scaling behavior of the model. In the quantum sector our model is related to the strong regime of the $t$-$U$ Hubbard model introduced in \cite{chico} and solved with diagonal open boundary condition. Although the open chain $t$-$U$ Hubbard model has a scaling gap of $L^{-1}$, since its spectrum coincides with the spectrum of the anisotropic XXZ model \cite{chico} at $\Delta=0$, our model with periodic boundary condition displays a scaling gap of $L^{-3}$.


\section{The asymmetric exclusion model with $N-1$ kinds of impurities}

We generalize the model discussed in the previous section by adding more types of impurities. Although we can formulate a general model including both stochastic process and quantum spin chains, we will consider for simplicity only stochastic process. The model introduced describes the dynamics of $N$ types of particles on an one-dimensional lattice of $L$ sites, where the total number $n_1$, $n_2$, ..., $n_N$ of particles of each type is conserved. Different from the case with just one kind of impurity \cite{LazoAnd}, discussed in the previous section, in our generalized model we can have more than one particle on each site (multiple site occupation). In order to describe the occupancy of a given site $i$ ($i=1,2,...,L$) we attach on site $i$ a set $\{\alpha\}_i=\{\alpha_1,...,\alpha_n\}$, where $\alpha_j=1,...,N$ ($j=1,...,n$) denotes a particle of kind $\alpha_j$. If $\{\alpha\}_i=\emptyset$, the site is vacant. If $\{\alpha\}_i=\{\alpha_1,...,\alpha_n\}$, we have on the site $n$ particles of kinds $\alpha_1$, $\alpha_2$,..., $\alpha_n$ with $\alpha_{j+1}>\alpha_j+1$ ($j=1,...,n-1$). The allowed configurations, denoted by $\lbrace\alpha\rbrace=\{ \{\alpha\}_1,\{\alpha\}_2,...,\{\alpha\}_L\}$ are those satisfying the hard-core constraints imposed by the condition $\alpha_{j+1}>\alpha_j+1$ for particles on the same site (see Figure \rf{fig1} for an example of an allowed configuration).

\begin{figure}[ht]
\includegraphics[width=\textwidth]{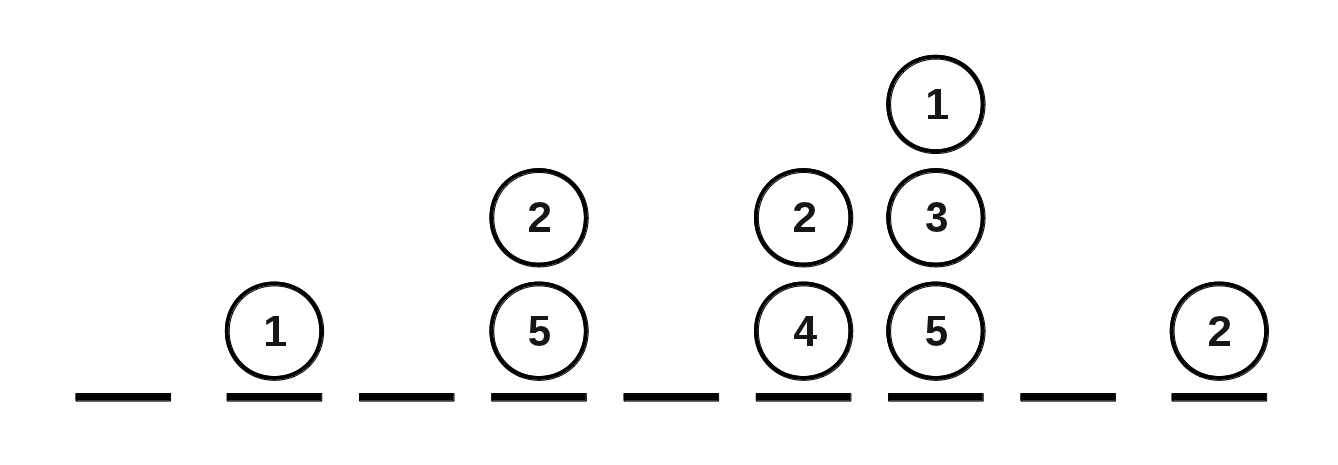}
\caption{In this figure we have an allowed particles configuration for $L=9$, $N=5$ and $n=n_1+n_2+n_3+n_4+n_5=9$. The configuration satisfy the hard-core constraints imposed by the condition $\alpha_{j+1}>\alpha_j+1$.}\lb{fig1}
\end{figure} 

The master equation for the probability distribution at a given time $t$, $P(\lbrace\alpha\rbrace,t)$, can be written in general as
\beq
\label{f1} 
\frac{\partial P(\lbrace\alpha\rbrace,t)}{\partial t} = - \Gamma(\lbrace\alpha\rbrace 
\rightarrow \lbrace\alpha'\rbrace) P(\lbrace\alpha\rbrace,t) + \Gamma(\lbrace\alpha'\rbrace 
\rightarrow \lbrace\alpha\rbrace) P(\lbrace\alpha'\rbrace,t)
\eeq
where $\Gamma(\lbrace\alpha\rbrace \rightarrow \lbrace\alpha'\rbrace)$ is 
the transition rate where a configuration $\lbrace\alpha\rbrace$ changes to $\lbrace\alpha'\rbrace$. In the model proposed there are only diffusion processes. As in \cite{LazoAnd}, and in the last section, if the neighbor sites are empty, particles of type $\alpha=1$ can jump to the right or to the left with rate $\Gamma_{0\;1}^{1\;0}$ and $\Gamma_{1\;0}^{0\;1}$, respectively. Particles of type $\alpha=2,...,N$ (impurities) do not jump to the neighbor sites if they are empty. The only allowed motions for impurities are those in which $2l$ ($l=1,2,...$) particles exchange positions (when $\alpha_{l+1}>2l+1$)
\bea
\label{f2a}
 &&\{1,3,...,2l-1,\alpha_{l+1},...\}_i\{2,4,...,2l,\beta_{l+1},...\}_{i+1} \\
&&\;\;\;\;\;\;\;\;\;\;\;\;\;\;\;\;\;\;\;\;\;\;\;\;\;\;\;\;\;\; \Rightarrow\{2,...,2l,\alpha_{l+1},...\}_i\{1,...,2l-1,\beta_{l+1},...\}_{i+1}\nonumber 
\eea
with transition rate $\Gamma^{1\;0}_{0\;1}$, and
\bea
\label{f2b}
 &&\{2,...,2l,\beta_{l+1},...\}_i\{1,...,2l-1,\alpha_{l+1},...\}_{i+1} \\
&&\;\;\;\;\;\;\;\;\;\;\;\;\;\;\;\;\;\;\;\;\;\;\;\;\;\;\;\;\;\; \Rightarrow\{1,3,...,2l-1,\beta_{l+1},...\}_i\{2,4,...,2l,\alpha_{l+1},...\}_{i+1}\nonumber 
\eea
with transition rate $\Gamma^{0\;1}_{1\;0}$, respectively, or those in which $2l+1$ particles exchange positions (when $\beta_{l+1}>2l+2$)
\bea
\label{f3a}
&&\{1,3,...,2l+1,\alpha_{l+2},...\}_i\{2,4,...,2l,\beta_{l+1},...\}_{i+1} \\ &&\;\;\;\;\;\;\;\;\;\;\;\;\;\;\;\;\;\;\;\;\;\;\;\;\;\;\;\;\;\; \Rightarrow \{2,...,2l,\alpha_{l+2},...\}_i\{1,...,2l+1,\beta_{l+1},...\}_{i+1}, \nonumber
\eea
with transition rates $\Gamma^{1\;0}_{0\;1}$, and
\bea
\label{f3b}
&&\{2,...,2l,\beta_{l+1},...\}_i\{1,...,2l+1,\alpha_{l+2},...\}_{i+1} \\ &&\;\;\;\;\;\;\;\;\;\;\;\;\;\;\;\;\;\;\;\;\;\;\;\;\;\;\;\;\;\; \Rightarrow \{1,3,...,2l+1,\beta_{l+1},...\}_i\{2,4,...,2l,\alpha_{l+2},...\}_{i+1}, \nonumber
\eea
with transition rates $\Gamma^{0\;1}_{1\;0}$, respectively. It is important to notice that all allowed motions are those in which we have a sequence of particles $1,3,...,2l\pm 1$ on site $i$ ($i+1$) and a sequence of particles $2,4,...,2l$ on site $i+1$ ($i$), respectively (see Figure \rf{fig2} for an example of an allowed motion of particles).
\begin{figure}[ht]
\includegraphics[width=\textwidth]{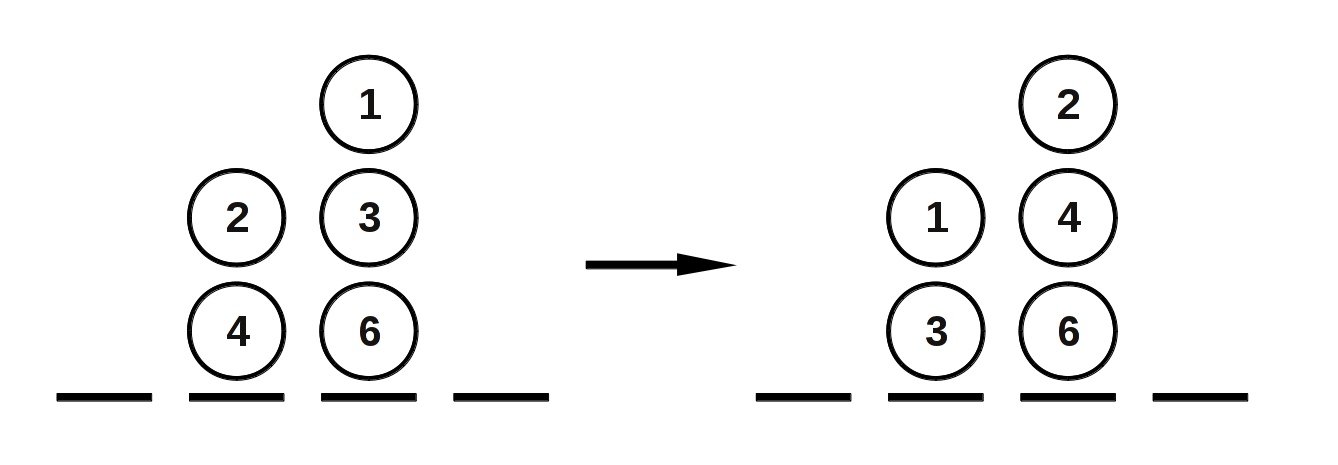} 
\caption{In this figure we have an allowed motion of four particles with transition rates $\Gamma^{1\;0}_{0\;1}$.}\lb{fig2}
\end{figure}
 
The stochastic Hamiltonian associated to the master equation \rf{f1} on a one-dimensional lattice of $L$ sites and periodic boundary condition is given by
\bea
\lb{f5}
H&=&\sum_{i=1}^L \left[\Gamma_{0\;1}^{1\;0}( E_i^{0,1}E_{i+1}^{1,0}- E_i^{1,1}E_{i+1}^{0,0})+\Gamma_{1\;0}^{0\;1}( E_i^{1,0}E_{i+1}^{0,1}-E_i^{0,0}E_{i+1}^{1,1})\right.\\
&&\!\!\!+\sum_{\{\alpha\}_i,\{\beta\}_{i+1}} \left.\Gamma_{\beta'\;\alpha'}^{\alpha\;\beta}(E_i^{\{\beta'\}_i,\{\alpha\}_i}E_{i+1}^{\{\alpha'\}_{i+1},\{\beta\}_{i+1}}-E_i^{\{\alpha\}_i,\{\alpha\}_i}E_{i+1}^{\{\beta\}_{i+1},\{\beta\}_{i+1}})\right], \nonumber
\eea
where the summations $\{\alpha\}_i$ and $\{\beta\}_{i+1}$ extends over all possible configurations satisfying the hard-core constraints imposed by the condition $\alpha_{j+1}>\alpha_j+1$ and $\beta_{j+1}>\beta_j+1$ for particles on the same site, and $E_i^{\{\alpha\}_i,\{\beta\}_i}$ (with $E^{\{\alpha\}_{L+1},\{\beta\}_{L+1}}_{L+1} \equiv E^{\{\alpha\}_{1},\{\beta\}_{1}}_{1}$ due to the periodic boundary condition) are operators that annihilate the configuration $\{\beta\}_i$ and create a configuration $\{\alpha\}_i$ on site $i$. The coupling constants $\Gamma_{\beta'\;\alpha'}^{\alpha\;\beta}$ in \rf{f5} are equal to $\Gamma^{1\;0}_{0\;1}$ or $\Gamma^{0\;1}_{1\;0}$ if $\{\alpha\}_i\;\{\beta\}_{i+1}\rightarrow \{\beta'\}_i\;\{\alpha'\}_{i+1}$ satisfy the relations \rf{f2a} and \rf{f2b}, or \rf{f3a} and \rf{f3b}, respectively, and are zero otherwise.

\subsection{The exact solution for the asymmetric exclusion model with $N-1$ types of impurities}

As in the case of one kind of impurity discussed in the previous section, we want to formulate a matrix product ansatz for the eigenvectors $|\Psi_{n_1,n_2,...,n_N,P}\rangle$ of the eigenvalue equation
\beq
H|\Psi_{n_1,n_2,...,n_N,P}\rangle=\varepsilon^{n_1,n_2,...,n_N}|\Psi_{n_1,n_2,...,n_N,P}\rangle
\lb{f5b}
\eeq
belonging to the eigensector labeled by ($n_1,n_2,...,n_N,P$). The MPA we propose asserts that the amplitudes corresponding to the configurations where there are no multiple occupancy are given by
\bea
\lb{f6}
&&f(x_1,\alpha_1;\ldots;x_n,\alpha_n) \Longleftrightarrow \\
&&\;\;\;\;\;\;\;\;\;\;  E^{x_1-1}A^{(\alpha_1)}E^{x_2-x_1-1}A^{(\alpha_2)}\cdots E^{x_n-x_{n-1}-1}A^{(\alpha_n)}E^{L-x_n},\nonumber
\eea
while if there exists a multiple occupancy with $m$ particles $\beta_1,\beta_2,...,\beta_m$ ($\beta_{i+1}>\beta_i+1$) at $x_j$ we have
\bea
\lb{f7}
&&f(\ldots;x_j,\beta_1;x_j,\beta_2;...;x_j,\beta_m;\ldots)\Longleftrightarrow\\
&&\;\;\;\;\;\;\;\;\;\; E^{x_1-1}A^{(\alpha_1)} \cdots E^{x_j-x_{j-1}-1}B^{(\beta_1,...,\beta_m)} E^{x_{j+m}-x_j-1} \cdots A^{(\alpha_n)}E^{L-x_n},\nonumber
\eea
where the matrices $A^{(\alpha)}$ are associated to the particles of type $\alpha$ ($\alpha=1,2,3$), the matrices $B^{(\beta_1,...,\beta_m)}=A^{(\beta_1)}E^{-1}A^{(\beta_2)}E^{-1}\cdots A^{(\beta_m)}$ are associated to a multiple occupation of $m$ particles $\beta_1,\beta_2,...,\beta_m$ at same site, and the matrix $E$ is associated to vacant sites. Furthermore, as the Hamiltonian \rf{f5} commutes with the momentum operator due to the periodic boundary condition, the amplitudes $f(x_1,\alpha_1;\ldots;x_n,\alpha_n)$ should satisfy the following relations:
\beq
\lb{f7a1}
f(x_1,\alpha_1;\ldots;x_n,\alpha_n)=e^{-iP}f(x_1+1,\alpha_1;\ldots;x_n+1,\alpha_n),
\eeq
where
\beq
\lb{f7a2}
P=\frac{2\pi l}{L}, \; l=0,1,...,L-1.
\eeq

The eigenvalue equation \rf{f5b} give us two kinds of relations for the amplitudes \rf{f6} and \rf{f7}. The first kind is related to those amplitudes without multiple occupancy, and the second type is related to those amplitudes with multiple occupancy. Let us consider separated each case:

{\bf Without multiple occupancy:} In this case, we have from the eigenvalue equation \rf{f5b} relations for amplitudes without collisions (particles of kind $1$ have only empty neighboring sites) and with collisions. For configuration without collision the amplitudes should satisfy the following constraints:
\bea
\lb{f7b}
&&\varepsilon^{n_1,n_2,...,n_N} f(x_1,\alpha_1;...;x_n,\alpha_n) \\
&&\;\;\;\;\;\;\;=\sum_{i=1}^n \left[\Gamma^{\alpha_i\;0}_{0\;\alpha_i}f(...;x_i-1,\alpha_i;...)+\Gamma_{\alpha_i\;0}^{0\;\alpha_i}f(...;x_i+1,\alpha_i;...)\right] \nonumber\\
&&\;\;\;\;\;\;\;\;\;\;-n_1f(x_1,\alpha_1;...;x_n,\alpha_n),\nonumber
\eea
where we introduced $\Gamma_{\alpha\;0}^{0\;\alpha}=\Gamma^{\alpha\;0}_{0\;\alpha}=0$ for $\alpha \neq 1$. In order to obtain a solution for \rf{f7b} we need to generalize \rf{e12} for $N$ kinds of particles. We consider the matrices $A^{(\alpha)}$ ($\alpha=1,...,N$) as composed by $n_{\alpha}$ spectral parameter dependent matrices $A_{k_j^{(\alpha)}}^{(\alpha)}$ belonging to $N$ distinct sets of spectral parameters, i. e.,
\beq
A^{(\alpha)}=\sum_{j=1}^{n_{\alpha}} EA_{k_j^{(\alpha)}}^{(\alpha)} \;\;\; \mbox{with} \;\;\; EA_{k_j^{(\alpha)}}^{(\alpha)}=e^{ik_j^{(\alpha)}}A_{k_j^{(\alpha)}}^{(\alpha)}E, \;\;\; \left(A_{k_j^{(\alpha)}}^{(\alpha)} \right)^2=0,
\lb{f8}
\eeq
for $\alpha=1,2,...,N$ and $n_1+n_2+\cdots +n_N=n$. These expressions when inserted into relations without collisions \rf{f7b} give us the energy in terms of the spectral parameters $\{k_j^{(1)}\}$
\beq
\varepsilon^{n_1,n_2,...,n_N}=\sum_{j=1}^{n_1}\varepsilon(k_j^{(1)}),
\lb{f9}
\eeq
where
\beq
\varepsilon(k)=\Gamma^{1\;0}_{0\;1}e^{-ik}+\Gamma_{1\;0}^{0\;1}e^{ik}-1.
\lb{f10}
\eeq
It is important to notice that, like in the previous section, the eigenvalues of the model depend only on the spectral parameters of particles of kind $1$. On the other hand, for amplitudes without multiple occupancy and particles of kind $\alpha$ and $\beta$ at the colliding positions ($x_{j+1}=x_j+1$), the eigenvalue equation \rf{f5b} give us the generalizations of \rf{e30}
with $\alpha,\beta=1,2$, where the $S$-matrix elements are given by
\bea
\lb{f10b}
S_{1\;1}^{1\;1}(k_j^{(1)},k_l^{(1)})&=&-\frac{\Gamma_{0\;1}^{1\;0}+\Gamma_{1\;0}^{0\;1}e^{i(k_j^{(1)}+k_l^{(1)})}-e^{ik_j^{(1)}}}{\Gamma_{0\;1}^{1\;0}+\Gamma_{1\;0}^{0\;1}e^{i(k_j^{(1)}+k_l^{(1)})}-e^{ik_l^{(1)}}},\\
S^{2 \; 1}_{2 \; 1}(k^{(2)},k^{(1)})&=&\frac{1}{S^{1 \; 2}_{1 \; 2}(k^{(1)},k^{(2)})}=e^{ik^{(2)}}. \nonumber
\eea
We do not consider here the cases where we have one particle of kind $1$ and other particle of kind greater than $2$ since the eigenvalue equation in this case will relates amplitudes with multiple occupancy. For $\alpha,\beta\geq 2$ the relations coming from the eigenvalue equation are identically satisfied. In this case, as in the previous section, we can choose without loss of generality
\beq
\lb{f10c}
S_{\alpha\;\alpha}^{\alpha\;\alpha}(k_j^{(\alpha)},k_l^{(\alpha)})=1 \;\;\;\;\; (\alpha\geq 2).
\eeq
Furthermore, in order to have amplitudes with non null norm, the spectral parameters $\{ k_j^{(\alpha)}\}$ ($\alpha \geq 2$) should satisfy the constraints
\beq
k_l^{(\alpha)}\neq k_j^{(\alpha)} + \frac{\pi (2m+1)}{d_v^{(\alpha)}} \;\; (m=0,1,...) \;\; (\alpha \geq 2),
\lb{f10d}
\eeq
where $\{d_v^{(\alpha)}\}$ is the set of all numbers of vacant sites between particles of type $\alpha \geq 2$.

{\bf With multiple occupancy:} Let us consider first the relations coming from the eigenvalue equation \rf{f5b} where we have only one site with multiple occupancy of $m$ particles $\beta_1,\beta_2,...,\beta_m$ ($\beta_{i+1}>\beta_i+1$) at position $x_j\equiv x$. We have from the eigenvalue equation \rf{f5b} the following relations:
\bea
\lb{f11}
&&\varepsilon^{n_1,n_2,...,n_N} f(...;x,\beta_1;x,\beta_2;...;x,\beta_m;...) \\
&&\;\;\;\;\;\;\;=\sum_{i=1,x_i\neq x}^n \left[\Gamma^{\alpha_i\;0}_{0\;\alpha_i}f(...;x_i-1,\alpha_i;...)+\Gamma_{\alpha_i\;0}^{0\;\alpha_i}f(...;x_i+1,\alpha_i;...)\right] \nonumber\\
&&\;\;\;\;\;\;\;\;\;\;+\Gamma^{\beta_1\;0}_{0\;\beta_1}f(...; x-1,\beta_1;x,\beta_2;...,x,\beta_m;...)\nonumber\\
&&\;\;\;\;\;\;\;\;\;\;+\Gamma_{\beta_1\;0}^{0\;\beta_1}f(...;x,\beta_2;...,x,\beta_m;x+1,\beta_1;...)\nonumber\\
&&\;\;\;\;\;\;\;\;\;\;-n_1f(...;x,\beta_1;x,\beta_2;...;x,\beta_m;...),\nonumber
\eea
for empty neighboring sites, and 
\bea
\lb{f12}
&&\varepsilon^{n_1,n_2,...,n_N} f(...;x-1,\beta_1;x,\beta_2;...;x,\beta_m;...) \\
&&\;\;\;\;\;\;\;=\sum_{i=1,x_i\neq x-1,x}^n \left[\Gamma^{\alpha_i\;0}_{0\;\alpha_i}f(...;x_i-1,\alpha_i;...)+\Gamma_{\alpha_i\;0}^{0\;\alpha_i}f(...;x_i+1,\alpha_i;...)\right] \nonumber\\
&&\;\;\;\;\;\;\;\;\;\;+\Gamma^{\beta_1\;0}_{0\;\beta_1}f(...; x-2,\beta_1;x,\beta_2;...,x,\beta_m;...)\nonumber\\
&&\;\;\;\;\;\;\;\;\;\;+\Gamma_{\beta_1\;0}^{0\;\beta_1}f(...;x,\beta_1;x,\beta_2;...,x,\beta_m;...)\nonumber\\
&&\;\;\;\;\;\;\;\;\;\;-n_1f(...;x,\beta_1;x,\beta_2;...;x,\beta_m;...),\nonumber
\eea
and
\bea
\lb{f13}
&&\varepsilon^{n_1,n_2,...,n_N} f(...;x,\beta_2;...;x,\beta_m;x+1,\beta_1...) \\
&&\;\;\;\;\;\;\;=\sum_{i=1,x_i\neq x,x+1}^n \left[\Gamma^{\alpha_i\;0}_{0\;\alpha_i}f(...;x_i-1,\alpha_i;...)+\Gamma_{\alpha_i\;0}^{0\;\alpha_i}f(...;x_i+1,\alpha_i;...)\right] \nonumber\\
&&\;\;\;\;\;\;\;\;\;\;+\Gamma^{\beta_1\;0}_{0\;\beta_1}f(...;x,\beta_1;x,\beta_2;...,x,\beta_m;...)\nonumber\\
&&\;\;\;\;\;\;\;\;\;\;+\Gamma_{\beta_1\;0}^{0\;\beta_1}f(...;x,\beta_2;...,x,\beta_m;x+2,\beta_1;...)\nonumber\\
&&\;\;\;\;\;\;\;\;\;\;-n_1f(...;x,\beta_1;x,\beta_2;...;x,\beta_m;...),\nonumber
\eea
for neighboring sites occupied by one particle of kind $\beta_1$.  In \rf{f11}, \rf{f12} and \rf{f13} we have $\beta_1=1,2,...,N$ and $\Gamma^{\beta_1\;0}_{0\;\beta_1}=\Gamma_{\beta_1\;0}^{0\;\beta_1}=0$ if $\beta_1\neq 1$, and without loss of generality we also choose no collisions of particles $1$. Equations \rf{f11}, \rf{f12} and \rf{f13} are automatically satisfied if $\beta_1\neq 1$. On the other hand, for $\beta_1=1$, while \rf{f12} is again automatically satisfied, the equations \rf{f11} and \rf{f13} will impose algebraic constraints for the matrices defining the ansatz. By inserting the ansatz \rf{f7} with \rf{f8} and \rf{f9} into equations \rf{f11} and \rf{f13} we obtain, after some algebraic manipulations, the following constraints among the matrices:
\beq
A_{k^{(1)}}^{(1)}A_{k^{(\beta_2)}}^{(\beta_2)}\cdots A_{k^{(\beta_m)}}^{(\beta_m)}=A_{k^{(\beta_2)}}^{(\beta_2)}\cdots A_{k^{(\beta_m)}}^{(\beta_m)}A_{k^{(1)}}^{(1)},
\lb{f14}
\eeq
where $\beta_2=3,4,...,N$ and $m=2,3,...$ with $\beta_{i+1}>\beta_i+1$. In order to satisfy \rf{f14} the for any set $\{\beta_2,...,\beta_m\}$ we should impose the following commutation relations among matrices $A_{k_j^{(1)}}^{(1)}$ and $A_{k_j^{(\alpha)}}^{(\alpha)}$ ($\alpha \geq 3$):
\beq
A_{k_j^{(1)}}^{(1)}A_{k_j^{(\alpha)}}^{(\alpha)}=A_{k_j^{(\alpha)}}^{(\alpha)}A_{k_j^{(1)}}^{(1)} \;\;\;\;\; (\alpha \geq 3).
\lb{f15}
\eeq

Let us consider now the configurations where we have neighbors sites at positions $x$ and $x+1$ with multiple particle occupations. For those configurations in which $2l$ ($l = 1, 2, ...$) particles exchange positions the eigenvalue equation \rf{f5b} give us the following relations:
\bea
\lb{f16}
&&\varepsilon^{n_1,n_2,...,n_N} f(...;x,1;...;x,2l-1;x,\beta_{l+1};...;x,\beta_m;\\
&&\;\;\;\;\;\;\;\;\;\;\;\;\;\;\;\;\;\;\;\;\;\;\;\;\;\;\;\;\;\;x+1,2;...;x+1,2l;x+1,\beta'_{l+1};...;x+1,\beta'_{m'};...)\nonumber \\
&&\;\;\;\;\;\;\;=\sum_{i=1,x_i\neq x,x+1}^n \left[\Gamma^{\alpha_i\;0}_{0\;\alpha_i}f(...;x_i-1,\alpha_i;...)+\Gamma_{\alpha_i\;0}^{0\;\alpha_i}f(...;x_i+1,\alpha_i;...)\right] \nonumber\\
&&\;\;\;\;\;\;\;\;\;\;+\Gamma^{1\;0}_{0\;1}f(...; x-1,1;x,3;...;x,\beta_m;...)\nonumber\\
&&\;\;\;\;\;\;\;\;\;\;+\Gamma^{0\;1}_{1\;0}f(...;x,2;...;x,2l;x,\beta_{l+1};...;x,\beta_m;\nonumber\\
&&\;\;\;\;\;\;\;\;\;\;\;\;\;\;\;\;\;\;\;\;\;\;\;\;\;\;\;\;\;\;x+1,1;...;x+1,2l-1;x+1,\beta'_{l+1};...;x+1,\beta'_{m'};...)\nonumber\\
&&\;\;\;\;\;\;\;\;\;\;-n_1f(...;x,1;...;x,2l-1;x,\beta_{l+1};...;x,\beta_m;\nonumber\\
&&\;\;\;\;\;\;\;\;\;\;\;\;\;\;\;\;\;\;\;\;\;\;\;\;\;\;\;\;\;\;x+1,2;...;x+1,2l;x+1,\beta'_{l+1};...;x+1,\beta'_{m'};...)\nonumber
\eea
and
\bea
\lb{f17}
&&\varepsilon^{n_1,n_2,...,n_N} f(...;x,2;...;x,2l;x,\beta'_{l+1};...;x,\beta'_{m'};\\
&&\;\;\;\;\;\;\;\;\;\;\;\;\;\;\;\;\;\;\;\;\;\;\;\;\;\;\;\;\;\;x+1,1;...;x+1,2l-1;x+1,\beta_{l+1};...;x+1,\beta_{m};...)\nonumber \\
&&\;\;\;\;\;\;\;=\sum_{i=1,x_i\neq x,x+1}^n \left[\Gamma^{\alpha_i\;0}_{0\;\alpha_i}f(...;x_i-1,\alpha_i;...)+\Gamma_{\alpha_i\;0}^{0\;\alpha_i}f(...;x_i+1,\alpha_i;...)\right] \nonumber\\
&&\;\;\;\;\;\;\;\;\;\;+\Gamma_{1\;0}^{0\;1}f(...; x+1,3;...;x+1,\beta'_{m'};x+2,1;...)\nonumber\\
&&\;\;\;\;\;\;\;\;\;\;+\Gamma_{0\;1}^{1\;0}f(...;x,1;...;x,2l-1;x,\beta'_{l+1};...;x,\beta'_{m'};\nonumber\\
&&\;\;\;\;\;\;\;\;\;\;\;\;\;\;\;\;\;\;\;\;\;\;\;\;\;\;\;\;\;\;x+1,2;...;x+1,2l;x+1,\beta_{l+1};...;x+1,\beta_{m};...)\nonumber\\
&&\;\;\;\;\;\;\;\;\;\;-n_1f(...;x,2;...;x,2l;x,\beta'_{l+1};...;x,\beta'_{m'};\nonumber\\
&&\;\;\;\;\;\;\;\;\;\;\;\;\;\;\;\;\;\;\;\;\;\;\;\;\;\;\;\;\;\;x+1,1;...;x+1,2l-1;x+1,\beta_{l+1};...;x+1,\beta_{m};...),\nonumber
\eea
where $\beta_{l+1}>2l+1$. Furthermore, for those configurations in which $2l+1$ ($l = 0,1, ...$) particles exchange positions the eigenvalue equation \rf{f5b} give us
\bea
\lb{f18}
&&\varepsilon^{n_1,n_2,...,n_N} f(...;x,1;...;x,2l+1;x,\beta_{l+2};...;x,\beta_m;\\
&&\;\;\;\;\;\;\;\;\;\;\;\;\;\;\;\;\;\;\;\;\;\;\;\;\;\;\;\;\;\;x+1,2;...;x+1,2l;x+1,\beta'_{l+1};...;x+1,\beta'_{m'};...)\nonumber \\
&&\;\;\;\;\;\;\;=\sum_{i=1,x_i\neq x,x+1}^n \left[\Gamma^{\alpha_i\;0}_{0\;\alpha_i}f(...;x_i-1,\alpha_i;...)+\Gamma_{\alpha_i\;0}^{0\;\alpha_i}f(...;x_i+1,\alpha_i;...)\right] \nonumber\\
&&\;\;\;\;\;\;\;\;\;\;+\Gamma^{1\;0}_{0\;1}f(...; x-1,1;x,3;...;x,\beta_m;...)\nonumber\\
&&\;\;\;\;\;\;\;\;\;\;+\Gamma^{0\;1}_{1\;0}f(...;x,2;...;x,2l;x,\beta_{l+2};...;x,\beta_m;\nonumber\\
&&\;\;\;\;\;\;\;\;\;\;\;\;\;\;\;\;\;\;\;\;\;\;\;\;\;\;\;\;\;\;x+1,1;...;x+1,2l+1;x+1,\beta'_{l+1};...;x+1,\beta'_{m'};...)\nonumber\\
&&\;\;\;\;\;\;\;\;\;\;-n_1f(...;x,1;...;x,2l+1;x,\beta_{l+2};...;x,\beta_m;\nonumber\\
&&\;\;\;\;\;\;\;\;\;\;\;\;\;\;\;\;\;\;\;\;\;\;\;\;\;\;\;\;\;\;x+1,2;...;x+1,2l;x+1,\beta'_{l+1};...;x+1,\beta'_{m'};...)\nonumber
\eea
and
\bea
\lb{f19}
&&\varepsilon^{n_1,n_2,...,n_N} f(...;x,2;...;x,2l;x,\beta'_{l+1};...;x,\beta'_{m'};\\
&&\;\;\;\;\;\;\;\;\;\;\;\;\;\;\;\;\;\;\;\;\;\;\;\;\;\;\;\;\;\;x+1,1;...;x+1,2l+1;x+1,\beta_{l+2};...;x+1,\beta_{m};...)\nonumber \\
&&\;\;\;\;\;\;\;=\sum_{i=1,x_i\neq x,x+1}^n \left[\Gamma^{\alpha_i\;0}_{0\;\alpha_i}f(...;x_i-1,\alpha_i;...)+\Gamma_{\alpha_i\;0}^{0\;\alpha_i}f(...;x_i+1,\alpha_i;...)\right] \nonumber\\
&&\;\;\;\;\;\;\;\;\;\;+\Gamma_{1\;0}^{0\;1}f(...; x+1,3;...;x+1,\beta_{m};x+2,1;...)\nonumber\\
&&\;\;\;\;\;\;\;\;\;\;+\Gamma_{0\;1}^{1\;0}f(...;x,1;...;x,2l+1;x,\beta'_{l+1};...;x,\beta'_{m'};\nonumber\\
&&\;\;\;\;\;\;\;\;\;\;\;\;\;\;\;\;\;\;\;\;\;\;\;\;\;\;\;\;\;\;x+1,2;...;x+1,2l;x+1,\beta_{l+2};...;x+1,\beta_{m};...)\nonumber\\
&&\;\;\;\;\;\;\;\;\;\;-n_1f(...;x,2;...;x,2l;x,\beta'_{l+1};...;x,\beta'_{m'};\nonumber\\
&&\;\;\;\;\;\;\;\;\;\;\;\;\;\;\;\;\;\;\;\;\;\;\;\;\;\;\;\;\;\;x+1,1;...;x+1,2l+1;x+1,\beta_{l+2};...;x+1,\beta_{m};...),\nonumber
\eea
where $\beta'_{l+1}>2l+2$. By inserting the ansatz \rf{f7} with \rf{f8}, \rf{f9} and \rf{f15} into equations \rf{f16}-\rf{f19} we obtain, after some algebraic manipulations, the following constraints among the matrices:
\bea
\lb{f20}
&&e^{i(k^{(2)}+\cdots+k^{(2l)})}A_{k^{(1)}}^{(1)}\cdots A_{k^{(2l-1)}}^{(2l-1)} A_{k^{(\beta_{l+1})}}^{(\beta_{l+1})}\cdots A_{k^{(\beta_m)}}^{(\beta_m)}A_{k^{(2)}}^{(2)}\cdots A_{k^{(2l)}}^{(2l)}=\\
&&\;\;\;\;\;\;\;\;\;\; e^{i(k^{(3)}+\cdots+k^{(2l-1)})}A_{k^{(2)}}^{(2)}\cdots A_{k^{(2l)}}^{(2l)}A_{k^{(\beta_{l+1})}}^{(\beta_{l+1})}\cdots A_{k^{(\beta_m)}}^{(\beta_m)}A_{k^{(1)}}^{(1)}\cdots A_{k^{(2l-1)}}^{(2l-1)},\nonumber
\eea
and
\bea
\lb{f21}
&&e^{i(k^{(2)}+\cdots+k^{(2l)})}A_{k^{(1)}}^{(1)}\cdots A_{k^{(2l+1)}}^{(2l+1)} A_{k^{(\beta_{l+1})}}^{(\beta_{l+1})}\cdots A_{k^{(\beta_m)}}^{(\beta_m)}A_{k^{(2)}}^{(2)}\cdots A_{k^{(2l)}}^{(2l)}=\\
&&\;\;\;\;\;\;\;\;\;\; e^{i(k^{(3)}+\cdots+k^{(2l+1)})}A_{k^{(2)}}^{(2)}\cdots A_{k^{(2l)}}^{(2l)}A_{k^{(\beta_{l+1})}}^{(\beta_{l+1})}\cdots A_{k^{(\beta_m)}}^{(\beta_m)}A_{k^{(1)}}^{(1)}\cdots A_{k^{(2l+1)}}^{(2l+1)}.\nonumber
\eea
In order to satisfy equations \rf{f20} and \rf{f21} for all $l=0,1,...$ and $m=2,3,...$ with $\beta_{i+1}>\beta_i+1$, the matrices defining the ansatz should satisfy the algebraic relations:
\bea
\lb{f22}
A_{k^{(\alpha+1)}}^{(\alpha+1)}A_{k^{(\alpha)}}^{(\alpha)}&=& e^{ik^{(\alpha+1)}}A_{k^{(\alpha)}}^{(\alpha)}A_{k^{(\alpha+1)}}^{(\alpha+1)} \;\;\; (\alpha=1,...,N-1)\\
A_{k^{(\alpha)}}^{(\alpha)}A_{k^{(\beta)}}^{(\beta)}&=& A_{k^{(\beta)}}^{(\beta)}A_{k^{(\alpha)}}^{(\alpha)} \;\;\; (\beta>\alpha+1).\nonumber
\eea
Equations \rf{f22} with \rf{f8}, \rf{f10b}, \rf{f10c} and \rf{f15} completely fix the commutation relations among the matrices defining the ansatz:
\bea
\lb{f23}
EA_{k_j^{(\alpha)}}^{(\alpha)}=e^{ik_j^{(\alpha)}}A_{k_j^{(\alpha)}}^{(\alpha)}E, \;\;\; \left(A_{k_j^{(\alpha)}}^{(\alpha)} \right)^2=0,\\
A_{k_j^{(\alpha)}}^{(\alpha)}A_{k_l^{(\beta)}}^{(\beta)}=S_{\alpha\;\beta}^{\alpha\;\beta}(k_j^{(\alpha)},k_l^{(\beta)})A_{k_l^{(\beta)}}^{(\beta)}A_{k_j^{(\alpha)}}^{(\alpha)}\;\;\; (\alpha,\beta=1,2,...,N),\nonumber
\eea
where the coupling constants $S_{\alpha\;\beta}^{\alpha\;\beta}(k_j^{(\alpha)},k_l^{(\beta)})$ are given by:
\bea
\lb{f24}
S_{1\;1}^{1\;1}(k_j^{(1)},k_l^{(1)})=-\frac{\Gamma_{0\;1}^{1\;0}+\Gamma_{1\;0}^{0\;1}e^{i(k_j^{(1)}+k_l^{(1)})}-e^{ik_j^{(1)}}}{\Gamma_{0\;1}^{1\;0}+\Gamma_{1\;0}^{0\;1}e^{i(k_j^{(1)}+k_l^{(1)})}-e^{ik_l^{(1)}}},\\
S_{\alpha\;\alpha}^{\alpha\;\alpha}(k_j^{(\alpha)},k_l^{(\alpha)})=1 \;\;\; (2 \leq \alpha \leq N),\nonumber\\
S_{\alpha+1\;\alpha}^{\alpha+1\;\alpha}(k_j^{(\alpha+1)},k_l^{(\alpha)})=\frac{1}{S_{\alpha\;\alpha+1}^{\alpha\;\alpha+1}(k_l^{(\alpha)},k_j^{(\alpha+1)})}=e^{ik_j^{(\alpha+1)}}\;\;\; (1 \leq \alpha \leq N-1),\nonumber\\
S_{\alpha\;\beta}^{\alpha\;\beta}(k_j^{(\alpha)},k_l^{(\beta)})=S_{\beta\;\alpha}^{\beta\;\alpha}(k_l^{(\beta)},k_j^{(\alpha)})=1 \;\;\; (\alpha=1,...,N-1,\alpha+1<\beta\leq N).\nonumber
\eea
Finally, all other relations coming from the eigenvalue equation \rf{f5b} containing amplitudes with arbitrary number of particles on neighbors sites are automatically satisfied by the ansatz \rf{f6} and \rf{f7} with \rf{f8} and the algebraic relations \rf{f23}. Furthermore, the associativity of the algebra \rf{f23} provides a well-defined value for any product of matrices and it follows from the fact that the algebra \rf{f23} is diagonal and the structure constants \rf{f24} are $c$-numbers with the property $S_{\alpha\;\beta}^{\alpha\;\beta}(k_j^{(\alpha)},k_l^{(\beta)})S_{\beta\;\alpha}^{\beta\;\alpha}(k_l^{(\beta)},k_j^{(\alpha)})=1$ ($\alpha,\beta=1,...,N$).

In order to complete our solution we should fix the spectral parameters $\{ k_j^{(\alpha)}\}$ ($\alpha=1,...,N$). Like in previous section, the algebraic expression in \rf{f23} assures that any matrix product defining our ansatz can be expressed in terms of a simple matrix product $A_{k_1^{(1)}}^{(1)}\cdots  A_{k_{n_1}^{(1)}}^{(1)}A_{k_1^{(2)}}^{(2)}\cdots A_{k_{n_2}^{(2)}}^{(2)}\cdots A_{k_1^{(N)}}^{(N)}\cdots A_{k_{n_N}^{(N)}}^{(N)} E^L$. From the periodic boundary condition we obtain:
\bea
\lb{f25}
A_{k_1^{(1)}}^{(1)}\cdots A_{k_{j-1}^{(1)}}^{(1)}A_{k_{j}^{(1)}}^{(1)}A_{k_{j+1}^{(1)}}^{(1)} \cdots A_{k_{n_N}^{(N)}}^{(N)} E^L \\
=\prod_{l>j}^{n_1}S_{1\;1}^{1\;1}(k_j^{(1)},k_l^{(1)})e^{-i\sum_{q=1}^{n_2}k_{q}^{(2)}}e^{-ik_j^{(1)}L}A_{k_1^{(1)}}^{(1)}\cdots A_{k_{j-1}^{(1)}}^{(1)}A_{k_{j+1}^{(1)}}^{(1)} \cdots A_{k_{n_N}^{(N)}}^{(N)}E^LA_{k_{j}^{(1)}}^{(1)} \nonumber \\
=\prod_{l>j}^{n_1}S_{1\;1}^{1\;1}(k_j^{(1)},k_l^{(1)})e^{-i\sum_{q=1}^{n_2}k_{q}^{(2)}}e^{-ik_j^{(1)}L}A_{k_{j}^{(1)}}^{(1)}A_{k_1^{(1)}}^{(1)}\cdots A_{k_{j-1}^{(1)}}^{(1)}A_{k_{j+1}^{(1)}}^{(1)} \cdots A_{k_{n_N}^{(N)}}^{(N)} E^L\nonumber
\\
=-\prod_{l=1}^{n_1}S_{1\;1}^{1\;1}(k_j^{(1)},k_l^{(1)})e^{-i\sum_{q=1}^{n_2}k_{q}^{(2)}}e^{-ik_j^{(1)}L} \nonumber \\
\;\;\;\;\;\;\;\;\;\;\;\;\;\;\;\;\;\;\;\;\;\;\;\;\;\;\;\;\;\;\;\;\;\;\;\;\;\;\;\;\times A_{k_1^{(1)}}^{(1)}\cdots A_{k_{j-1}^{(1)}}^{(1)}A_{k_{j}^{(1)}}^{(1)}A_{k_{j+1}^{(1)}}^{(1)}
\cdots A_{k_{n_N}^{(N)}}^{(N)} E^L,\nonumber
\eea
and
\bea
\lb{f26}
A_{k_1^{(1)}}^{(1)}\cdots A_{k_{j-1}^{(\alpha)}}^{(\alpha)}A_{k_{j}^{(\alpha)}}^{(\alpha)}A_{k_{j+1}^{(\alpha)}}^{(\alpha)} \cdots A_{k_{n_N}^{(N)}}^{(N)} E^L \\
=e^{-i\sum_{q=1}^{n_{\alpha+1}}k_{q}^{(\alpha+1)}}e^{-ik_j^{(\alpha)}L}A_{k_1^{(1)}}^{(1)}\cdots A_{k_{j-1}^{(\alpha)}}^{(\alpha)}A_{k_{j+1}^{(\alpha)}}^{(\alpha)} \cdots A_{k_{n_N}^{(N)}}^{(N)}E^LA_{k_{j}^{(\alpha)}}^{(\alpha)} \nonumber \\
=e^{-i\sum_{q=1}^{n_{\alpha+1}}k_{q}^{(\alpha+1)}}e^{-ik_j^{(\alpha)}L}A_{k_{j}^{(\alpha)}}^{(\alpha)}A_{k_1^{(1)}}^{(1)}\cdots A_{k_{j-1}^{(\alpha)}}^{(\alpha)}A_{k_{j+1}^{(\alpha)}}^{(\alpha)} \cdots A_{k_{n_N}^{(N)}}^{(N)}E^L \nonumber \\
=e^{-i\sum_{q=1}^{n_{\alpha+1}}k_{q}^{(\alpha+1)}}e^{-ik_j^{(\alpha)}(L-n_{\alpha-1})}A_{k_1^{(1)}}^{(1)}\cdots A_{k_{j-1}^{(\alpha)}}^{(\alpha)}A_{k_{j}^{(\alpha)}}^{(\alpha)}A_{k_{j+1}^{(\alpha)}}^{(\alpha)} \cdots A_{k_{n_N}^{(N)}}^{(N)}E^L, \nonumber
\eea
for $\alpha=2,...,N-1$, and 
\bea
\lb{f27}
A_{k_1^{(1)}}^{(1)}\cdots A_{k_{j-1}^{(N)}}^{(N)}A_{k_{j}^{(N)}}^{(N)}A_{k_{j+1}^{(N)}}^{(N)} \cdots A_{k_{n_N}^{(N)}}^{(N)} E^L \\
=e^{-ik_j^{(N)}L}A_{k_1^{(1)}}^{(1)}\cdots A_{k_{j-1}^{(N)}}^{(N)}A_{k_{j+1}^{(N)}}^{(N)} \cdots A_{k_{n_N}^{(N)}}^{(N)}E^LA_{k_{j}^{(N)}}^{(N)} \nonumber \\
=e^{-ik_j^{(N)}L}A_{k_{j}^{(N)}}^{(N)}A_{k_1^{(1)}}^{(1)}\cdots A_{k_{j-1}^{(N)}}^{(N)}A_{k_{j+1}^{(N)}}^{(N)} \cdots A_{k_{n_N}^{(N)}}^{(N)}E^L \nonumber \\
=e^{-ik_j^{(N)}(L-n_{N-1})}A_{k_1^{(1)}}^{(1)}\cdots A_{k_{j-1}^{(N)}}^{(N)}A_{k_{j}^{(N)}}^{(N)}A_{k_{j+1}^{(N)}}^{(N)} \cdots A_{k_{n_N}^{(N)}}^{(N)}E^L, \nonumber
\eea
where in \rf{f25}, \rf{f25} and \rf{f27} we used the algebraic relations \rf{f23} with \rf{f24} and we introduced $S_{1\;1}^{1\;1}(k_j^{(1)},k_j^{(1)})=-1$. From \rf{f25}, \rf{f25} and \rf{f27} we obtain the Bethe equations for our model:
\bea
\lb{f28}
e^{ik_j^{(1)}L}=(-)^{n_1-1}e^{-i\sum_{q=1}^{n_2}k_{q}^{(2)}} \prod_{l=1}^{n_1}\frac{\Gamma_{0\;1}^{1\;0}+\Gamma_{1\;0}^{0\;1}e^{i(k_j^{(1)}+k_l^{(1)})}-e^{ik_j^{(1)}}}{\Gamma_{0\;1}^{1\;0}+\Gamma_{1\;0}^{0\;1}e^{i(k_j^{(1)}+k_l^{(1)})}-e^{ik_l^{(1)}}},\\
\lb{f29}
e^{ik_j^{(\alpha)}(L-n_{\alpha-1})}=e^{-i\sum_{q=1}^{n_{\alpha+1}}k_{q}^{(\alpha+1)}} \;\;\; (\alpha=2,...,N-1),\\
\lb{f30}
e^{ik_j^{(N)}(L-n_{N-1})}=1,
\eea
where equations should be satisfied for all $k_j^{(\alpha)}$ ($\alpha=1,...,N$) with $j=1,...,n_{\alpha}$. On the other hand, the momentum of the eigenstate is given by inserting the ansatz \rf{f6} and \rf{f7} into relation \rf{f7a1} and \rf{f7a2}:
\beq
\lb{f34}
P=\sum_{j=1}^{n_1} k_j^{(1)} + \sum_{j=1}^{n_2} k_j^{(2)}+\cdots + \sum_{j=1}^{n_N} k_j^{(N)}=\frac{2\pi l}{L} \;\;\; (l=0,1,\ldots,L-1).
\eeq
where the spectral parameters satisfy the Bethe equation \rf{f28}, \rf{f29} and \rf{f30}.

The Bethe equations \rf{f28}, \rf{f29} and \rf{f30} are more complicated than the case of previous section since the spectral parameters, and the eigenvalues of our model, depend on the densities of impurities. We compare the Bethe equation solution \rf{f33} for $N=3$ with the eigenvalues obtained from direct diagonalization of the Hamiltonian \rf{f5} with $L=5$, $n_1=2$, $n_2=1$ and $n_3=1$ (see Appendix B). The model displays the full spectrum of the ASEP and additional eigenvalues. The spectrum of the ASEP is obtained when $\sum_{j=1}^{n_{\alpha}}k_{j}^{(\alpha)}= 0$ for all $\alpha=2,...,N$. The stationary state belongs to this case and has the eigenvalue with the largest real part $\varepsilon^{n_1,n_2}=0$. In order to obtain the second largest real part eigenvalue, we can rewritten the Bethe equations in a more convenient way by eliminating the the spectral parameters $k_{j}^{(2)}$ in equation \rf{f28}. The first excited state is obtained when we have
\beq
\lb{f31}
\sum_{j=1}^{n_{\alpha}}k_{j}^{(\alpha)}=\frac{2\pi}{(L-n_{N-1})(L-n_{N-2})\cdots (L-n_{\alpha-1})}
\eeq
for all $\alpha=2,...,N$ (see Appendix B for a detailed discussion for a small system). Hence, by using \rf{f31} we can relate the sum over spectral parameters $k_{j}^{(2)}$ for the first excited state with roots of unity:
\beq
\lb{f32}
\sum_{q=1}^{n_2}k_{q}^{(2)}=\frac{2\pi}{L^{N-1}}\prod_{l=1}^{N-1}(1-\rho_l)^{-1},
\eeq
where $\rho_l=\frac{n_l}{L}$ are the densities of particles of kind $l=1,...,N$. Finally, inserting \rf{f32} into \rf{f28} we obtain the following Bethe equation:
\beq
\lb{f33}
e^{ik_j^{(1)}L}=(-)^{n_1-1}e^{-i\frac{2\pi }{L^{N-1}}\prod_{l=1}^{N-1}(1-\rho_l)^{-1}}\prod_{l=1}^{n_1}\frac{\Gamma_{0\;1}^{1\;0}+\Gamma^{0\;1}_{1\;0}e^{i(k_j^{(1)}+k_l^{(1)})}-e^{ik_j^{(1)}}}{\Gamma_{0\;1}^{1\;0}+\Gamma^{0\;1}_{1\;0}e^{i(k_j^{(1)}+k_l^{(1)})}- e^{ik_l^{(1)}}}.
\eeq

The Bethe equation \rf{f33} generalizes \cite{lazo2,LazoAnd} and \rf{e39} to the case of $N-1$ kinds of impurities. As in the case of one kind of impurities ($N=2$), the phase factor on \rf{f33} plays a fundamental role in the spectral properties of the model. The eigenvalue with the second largest real part determines the relaxation time and the dynamical exponent $z$. This eigenvalue is provided by selecting $n_1$ fugacities from \rf{f33} with momentum $P=\frac{2\pi}{L}$ in \rf{f34}. We solve \rf{f33} numerically for the totally asymmetric exclusion process (N-TASEPI), when $\Gamma_{0\;1}^{1\;0}=1$ and $\Gamma^{0\;1}_{1\;0}=0$ (or $\Gamma_{0\;1}^{1\;0}=0$ and $\Gamma^{0\;1}_{1\;0}=1$), in the half-filling sector $n_1=L/2$ for both two ($N=3$) and three ($N=4$) kinds of impurities (see Figure \rf{fig4}).
\begin{figure}[ht]
\includegraphics[width=\textwidth]{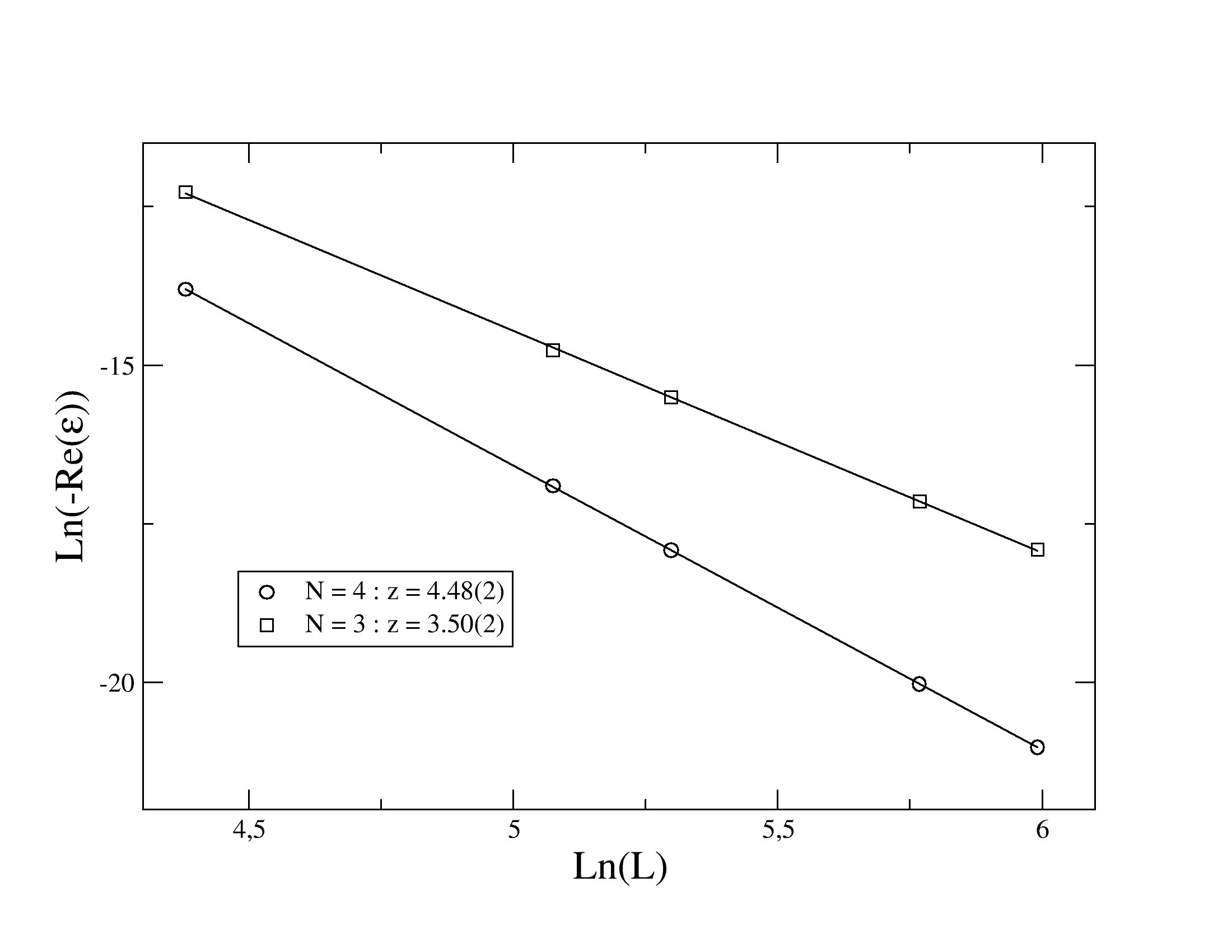}
\caption{In this figure we display the logarithm of the real part of the energy gap versus the logarithm of $L$ for $N=3,4$ in the in the half-filling sector $n_1=L/2$. The dynamical exponents $z$ are $3.50$ and $4.01(2)$, respectively.}\lb{fig4}
\end{figure} 
As in previous section, the errors displayed are computed from the linear regression for the logarithm of the real part of the energy gap versus the logarithm of $L=80,160,200,320,400$. For $N=3$ we consider $n_2=L/8$ and $n_3=L/8$. In this case we found that the spectral gap scales with $L^{-\frac{7}{2}}$ instead of $L^{-\frac{3}{2}}$ for the ASEP \cite{GwaSpohn} and $L^{-\frac{5}{2}}$ for the ASEPI \cite{LazoAnd}. On the other hand, for $N=4$ we choose $n_2=L/8$, $n_3=L/8$ and $n_4=L/8$, and we found that the spectral gap scales with $L^{-\frac{9}{2}}$. From these results, we can conjecture that for general $N=2,3,...$ our model will display a scaling exponent $z=\frac{3}{2}+N-1$. This scaling law is a consequence of the dependency of the dynamic of impurities of kind $N$ from the dynamic of impurities of kinds $\alpha<N$, since the impurities $\alpha$ only move when particles of kind $\alpha-1$ change position with them. As a consequence, the time to vanish the fluctuations on the densities of particles of kinds $N-1$ acts as a time scale for the diffusion of the impurities $N$ (reflected in the $L^{-\frac{3}{2}}\times L^{-N+1}$ gap for the N-ASEPI).

{\bf Eigenstates:} The stationary state is the eigenstate associated to the eigenvalue $\varepsilon^{n_1,n_2}=0$, and with momentum $P=0$. In this case, the Bethe equations \rf{f28}, \rf{f29} and \rf{f30} have the unique solution $e^{k_{j}^{(\alpha)}}=1$ for all $\alpha=1,...,N$ and $j=1,...,n_{\alpha}$, and the $S$-matrix reduces to the identity $S_{\alpha_1\;\alpha_2}^{\alpha_1\;\alpha_2}(k_j^{(\alpha_1)},k_l^{(\alpha_2)})=1$. Consequently, like in the previous section, at the stationary state all configurations that satisfy the hard-core constraints imposed by the definition of the model can occur with equal probabilities. Furthermore, each site is occupied by a particle $\alpha$ with probability $\rho_{\alpha}=n_{\alpha}/L$.


\section{Conclusion}

In the present work we formulate an exactly solvable asymmetrical diffusion model of $N=1,2,3,...$ kinds of particles with impurities (N-ASEPI). In this model particles of kind $1$ can jump to neighboring sites if they are empty and particles of kind $\alpha=2,3,...,N$ (called impurities) only exchange positions with others particles, satisfying a well defined dynamics. We solve this model with periodic boundary condition through a new matrix product ansatz \cite{alclazo,popkov} and we analyze the spectral gap for some special cases. Our N-ASEPI model displays the full spectrum of the ASEP \cite{GwaSpohn} plus new levels. The first excited state belongs to these new levels and has unusual scaling exponents. Although the ASEP belongs to the KPZ universality class, characterized by the dynamical exponent $z = \frac{3}{2}$ \cite{KPZ}, we conjecture that our model displays a scaling exponent $\frac{3}{2}+N-1$, where $N-1$ is the number of kinds of impurities. In order to check our conjecture, we solve numerically the Bethe equation with $N=3$ and $N=4$ for the totally asymmetric diffusion and we found that the gap for the first excited state scale as $L^{-\frac{7}{2}}$ and $L^{-\frac{9}{2}}$ in these cases. Furthermore, for $N=2$ we generalize the model \cite{LazoAnd} to include quantum spin chain Hamiltonians and we analyze the Bethe Ansatz equation for the symmetric and asymmetric diffusions. A quite interesting problem for the future concerns the formulation of the model with open boundary conditions instead periodic ones. In this case we expect the critical behavior of the model will display the same scaling exponent $\frac{3}{2}+N-1$.


\section*{Acknowledgments} 
 This work has been partly supported by CNPq, CAPES and FAPERGS (Brazilian agencies).


\appendix
\section{Eigenvalues and Bethe roots for $N=2$, $n_1=2$, $n_2=1$, and $L=6$}

In this appendix we list the full spectrum of the Hamiltonian \rf{e1} for the asymmetric exclusion model with one kind of impurity for $N=2$, $n_1=2$, $n_2=1$, $L=6$, and $\Gamma_{0 \; 1}^{1\; 0}=0.75$. We compared the eigenvalues obtained by direct diagonalization of the Hamiltonian \rf{e1} with those given by the Bethe ansatz solution. For $N=2$, $n_1=2$, $n_2=1$, $L=6$, and $\Gamma_{0 \; 1}^{1\; 0}=0.75$ the Bethe equations \rf{e41} reduces to
\beq
\lb{aA1}
x^6=-e^{-i\frac{2\pi}{4}m}\frac{0.75+0.25xy-x}{0.75+0.25xy- y},
\eeq
where $m=0,1,2,3$ and $x=e^{ik_1^{(1)}}$ and $y=e^{ik_2^{(1)}}$ are the fugacities. Equation \rf{aA1} can be reduced to a simple polynomial equation by inserting the relation for the momentum $P=\frac{2\pi l}{6}$ \rf{e35b}
\beq
\lb{aA2}
xy= e^{i\frac{2\pi l}{6}+i\frac{2\pi}{4}m},
\eeq
where $l=0,1,2,3,4,5$ and we have used \rf{e38}. In table \rf{table1} we display the full spectrum of the Hamiltonian \rf{e1} and the associated Bethe roots $x$ and $y$. The eigenvalue with the largest real part is zero, it is provided by choosing $m=0$ and $P=0$ ($l=0$) in \rf{aA1} and \rf{aA2}. In this case all spectral parameters are zero and we have $e^{ik_1^{(1)}}=e^{ik_2^{(1)}}=e^{ik_1^{(2)}}=1$. It is also important to note that for $m=0$ our model reproduces the full spectrum of the ASEP. For $m=1,2,3$ our model displays additionals energy levels. In special, the eigenvalues with the second largest real part, that determine the relaxation time and the dynamical exponent $z$, belong to these new levels. Actually, these eigenvalues are given by a complex conjugated pair by choosing $m=1$ and $P=\frac{2\pi}{L}=\frac{2\pi}{6}$ and by choosing $m=L-n_1-1=3$ and $P=(L-1)\frac{2\pi}{L}=5\frac{2\pi}{6}$.

\begin{table}[ht]
\caption{The eigenvalues for $L=6$, $n_1=2$ and $n_2=1$}\label{table1}
\vspace{3mm}
\centering
\begin{scriptsize}
\begin{tabular}{|c|c|c|}
\hline
Sector   & Energies & Bethe roots  \\ 
\hline
\multirow{3}{*}{$P=0$, $m=0$} & $0$ & $1$; $1$  \\ 
   & $-1.3819660$ & $0.3090170\pm 0.9510565i$ \\  
   & $-3.6180340$ & $-0.8090170\pm 0.5877852i$ \\
\hline
\multirow{2}{*}{$P=\frac{2\pi}{6}$, $m=0$} & $-0.5074856-0.2181393i$ & $0.8199801-0.0796133i$; $0.5024902+1.1049419i$  \\ 
   & $-2.4925144+0.2181393i$ & $-0.7655328+0.7087925i$; $0.2122933-0.9347130i$  \\  
\hline
\multirow{3}{*}{$P=2\frac{2\pi}{6}$, $m=0$} & $-0.8772838-0.5974379i$ & $0.5755876+0.3275559i$; $-0.0093985+1.5099420i$  \\ 
   & $-1.3562417-0.1208056i$ & $0.7592145-0.0749922i$; $-0.7637957+1.0652413i$ \\  
   & $-2.7664745+0.7182435i$ & $-0.9728436-0.4011708i$; $0.1255214-0.9419611i$ \\
\hline
\multirow{2}{*}{$P=3\frac{2\pi}{6}$, $m=0$} & $-1.5$ & $0.6180340$; $-1.6180340$  \\ 
   & $-1.5$ & $-1.6180340$; $0.6180340$ \\  
\hline
\multirow{3}{*}{$P=4\frac{2\pi}{6}$, $m=0$} & $-0.8772838+0.5974379i$ & $-0.0093985-1.5099420i$; $0.5755876-0.3275559i$  \\ 
   & $-1.3562417+0.1208056i$ & $0.7592145+0.0749922i$; $-0.7637957-1.0652413i$ \\  
   & $-2.7664745-0.7182435i$ & $0.1255214+0.9419611i$; $-0.9728436+0.4011708i$ \\
\hline
\multirow{2}{*}{$P=5\frac{2\pi}{6}$, $m=0$} & $-0.5074856+0.2181393i$ & $0.8199801+0.0796133i$; $0.5024902-1.1049419i$  \\ 
   & $-2.4925144-0.2181393i$ & $0.2122933+0.9347130i$; $-0.7655328-0.7087925i$ \\  
\hline
\multirow{2}{*}{$P=0$, $m=1$} & $-0.6557872+0.2961339i$ & $0.7018547-0.0959405i$; $0.1911910-1.3986613i$  \\ 
   & $-2.3442128-0.2961339i$ & $0.4489168+0.8140546i$; $-0.9419626-0.5194527i$ \\ 
\hline
\multirow{3}{*}{$P=\frac{2\pi}{6}$, $m=1$} & $-0.0572751+0.2059994i$ & $0.8373023-0.0855293i$; $1.0839923-0.4864275i$  \\ 
   & $-1.3828654+0.0130776i$ & $0.5293573+0.8049461i$; $0.0602956-1.0362278i$ \\  
   & $-3.5598595-0.2190769i$ & $-0.6226580+0.7713741i$; $-0.9411875-0.3629724i$ \\
\hline
\multirow{2}{*}{$P=2\frac{2\pi}{6}$, $m=1$} & $-0.4139462-0.1150956i$ & $0.6884251+0.8429321i$; $0.8591734-0.3257069i$  \\ 
   & $-2.5860538+0.1150956i$ & $-0.5524868+0.8570897i$; $-0.0480098-0.9794781i$  \\  
\hline
\multirow{3}{*}{$P=3\frac{2\pi}{6}$, $m=1$} & $-0.4977725-0.5315017i$ & $0.6345797+0.2963134i$; $0.6041134+1.2937589$  \\ 
   & $-1.3835508-0.0616583i$ & $0.7884227-0.3472311i$; $-0.4678531+1.0623069i$ \\  
   & $-3.1186767+0.5931600i$ & $-1.0219882-0.1463166i$; $-0.1372745-0.9588315i$ \\
\hline
\multirow{2}{*}{$P=4\frac{2\pi}{6}$, $m=1$} & $-1.0832361-0.2999300i$ & $0.4953389+0.1442693i$; $-1.3406309+1.3998737i$  \\ 
   & $-1.9167639+0.2999300i$ & $-1.2180653+0.0720208i$; $0.7326938-0.3671649i$ \\  
\hline
\multirow{3}{*}{$P=5\frac{2\pi}{6}$, $m=1$} & $-1.2103358+0.1188691i$ & $0.5489836-0.0641372i$; $-1.4512935-1.0803272i$  \\ 
   & $-1.4330978+0.6646123i$ & $0.6425791-0.4326605i$; $-0.5668353-1.1597751i$ \\  
   & $-2.3565664-0.7834814i$ & $-0.8735004+0.6433986i$; $0.3694030+0.8445026i$ \\
\hline
\multirow{3}{*}{$P=0$, $m=2$} & $-1.1746854$ & $0.4712748$; $-2.1219040$  \\ 
   & $-1.9126573-0.7734344i$ & $0.5609520+0.6692321i$; $-0.7356374+0.8776368i$ \\  
   & $-1.9126573+0.7734344i$ & $0.5609520-0.6692321i$; $-0.7356374-0.8776368i$ \\
\hline
\multirow{2}{*}{$P=\frac{2\pi}{6}$, $m=2$} & $-0.8494232+0.3327914i$ & $0.5692219-0.1648546i$; $-0.4038910-1.6383922i$  \\ 
   & $-2.1505768-0.3327914i$ & $0.6340571+0.6212665i$; $-1.0851023-0.3026347i$  \\  
\hline
\multirow{3}{*}{$P=2\frac{2\pi}{6}$, $m=2$} & $-0.2257571+0.3909776i$ & $0.7355889-0.2131056i$; $0.9417640-0.9044865i$  \\ 
   & $-1.3848207+0.0309866i$ & $0.6985648+0.5996111i$; $-0.2005790-1.0675545i$ \\  
   & $-3.3894222-0.4219643i$ & $-0.3934528+0.8990944i$; $-1.0126551-0.1129670i$ \\
\hline
\multirow{2}{*}{$P=3\frac{2\pi}{6}$, $m=2$} & $-0.3819660$ & $0.8090170+0.5877853i$; $0.8090170-0.5877853i$  \\ 
   & $-2.6180340$ & $-0.3090170+0.9510565i$; $-0.3090170+0.9510565i$ \\  
\hline
\multirow{3}{*}{$P=4\frac{2\pi}{6}$, $m=2$} & $-0.2257571-0.3909776i$ & $0.7355889+0.2131056i$; $0.9417640+0.9044865i$  \\ 
   & $-1.3848207-0.0309866i$ & $-0.2005790+1.0675545i$; $0.6985648-0.5996111i$ \\  
   & $-3.3894222+0.4219643i$ & $-1.0126551+0.1129670i$; $-0.3934528-0.8990944i$ \\
\hline
\multirow{2}{*}{$P=5\frac{2\pi}{6}$, $m=2$} & $-0.8494232-0.3327914i$ & $0.5692219+0.1648546i$; $-0.4038910+1.6383922i$  \\ 
   & $-2.1505768+0.3327914i$ & $0.6340571-0.6212665i$; $-1.0851023+0.3026347i$  \\  
\hline
\multirow{2}{*}{$P=0$, $m=3$} & $-0.6557872-0.2961339i$ & $0.7018547+0.0959405i$; $0.1911910+1.3986613i$  \\ 
   & $-2.3442128+0.2961339i$ & $0.4489168-0.8140546i$; $-0.9419626+0.5194527i$ \\ 
\hline
\multirow{3}{*}{$P=\frac{2\pi}{6}$, $m=3$} & $-1.2103358-0.1188691i$ & $0.5489836+0.0641372i$; $-1.4512935+1.0803272i$  \\ 
   & $-1.4330978-0.6646123i$ & $0.6425791+0.4326605i$; $-0.5668353+1.1597751i$ \\  
   & $-2.3565664+0.7834814i$ & $-0.8735004-0.6433986i$; $0.3694030-0.8445026i$ \\
\hline
\multirow{2}{*}{$P=2\frac{2\pi}{6}$, $m=3$} & $-1.0832361+0.2999300i$ & $0.4953389-0.1442693i$; $-1.3406309-1.3998737i$  \\ 
   & $-1.9167639-0.2999300i$ & $0.7326938+0.3671649i$; $-1.2180653-0.0720208i$  \\  
\hline
\multirow{3}{*}{$P=3\frac{2\pi}{6}$, $m=3$} & $-0.4977725+0.5315017i$ & $0.6345797-0.2963134i$; $0.6041134-1.2937589$  \\ 
   & $-1.3835508+0.0616583i$ & $0.7884227+0.3472311i$; $-0.4678531-1.0623069i$ \\  
   & $-3.1186767-0.5931600i$ & $-1.0219882+0.1463166i$; $-0.1372745+0.9588315i$ \\
\hline
\multirow{2}{*}{$P=4\frac{2\pi}{6}$, $m=3$} & $-0.4139462+0.1150956i$ & $0.8591734+0.3257069i$; $0.6884251-0.8429321i$  \\ 
   & $-2.5860538-0.1150956i$ & $-0.0480098+0.9794781i$; $-0.5524868-0.8570897i$ \\  
\hline
\multirow{3}{*}{$P=5\frac{2\pi}{6}$, $m=3$} & $-0.0572751-0.2059994i$ & $0.8373023+0.0855293i$; $1.0839923+0.4864275i$  \\ 
   & $-1.3828654-0.0130776i$ & $0.5293573-0.8049461i$; $0.0602956+1.0362278i$ \\  
   & $-3.5598595+0.2190769i$ & $-0.6226580-0.7713741i$; $-0.9411875+0.3629724i$ \\
\hline
\end{tabular}
\end{scriptsize}
\end{table}


\section{Eigenvalues and Bethe roots for $N=3$, $n_1=2$, $n_2=1$, $n_3=1$ and $L=5$}

In this appendix we list the full spectrum of the Hamiltonian \rf{f5} for the asymmetric exclusion model with one kind of impurities for $N=3$, $n_1=2$, $n_2=1$, $n_3=1$, $L=5$, and $\Gamma_{0 \; 1}^{1\; 0}=1$. We compared the eigenvalues obtained by direct diagonalization of the Hamiltonian \rf{f5} with those given by the Bethe ansatz solution. For $N=3$, $n_1=2$, $n_2=1$, $n_3=1$, $L=5$, and $\Gamma_{0 \; 1}^{1\; 0}=1$ the Bethe equations \rf{f28}, \rf{f29} and \rf{f30} reduce to
\bea
\lb{aB1}
x^5=-e^{-ik^{(2)}} \frac{1-x}{1-y},\\
\lb{aB2}
e^{i3k^{(2)}}=e^{-ik^{(3)}},\\
\lb{aB3}
e^{i4k^{(3)}}=1,
\eea
where $x=e^{ik_1^{(1)}}$ and $y=e^{ik_2^{(1)}}$ are the fugacities. Equation \rf{aB1} can be reduced to a simple polynomial equation by inserting the relation for the momentum $P=\frac{2\pi l}{6}$ \rf{e35b}
\beq
\lb{aB4}
xy= e^{i\frac{2\pi l}{5}-i\frac{2\pi}{6}m-i\frac{2\pi}{3}m'},
\eeq
where $l=0,1,2,3,4,5$, $m=0,1,2,3$, $m'=0,1,2$, and we have used from \rf{aB2} and \rf{aB3} the following relations:
\bea
\lb{aB5}
k^{(2)}=-\frac{2\pi}{12}m+\frac{2\pi}{3}m',\\
\lb{aB6}
k^{(3)}=\frac{2\pi}{4}m.
\eea
In tables \rf{table3} and \rf{table3b} we display the full spectrum of the Hamiltonian \rf{f5} and the associated Bethe roots $x$ and $y$. The eigenvalue with the largest real part is zero, it is provided by choosing $m=0$, $m'=0$ and $P=0$ ($l=0$) in \rf{aB1}, \rf{aB4} and \rf{aB5}. In this case all spectral parameters are zero and we have $e^{ik_1^{(1)}}=e^{ik_2^{(1)}}=e^{ik^{(2)}}=e^{ik^{(3)}}=1$. As in the case $N=2$, it is important to note that for $m=m'=0$ our model reproduces the full spectrum of the ASEP. If $m$ or $m'$ are not zero our model displays additionals energy levels. In special, the eigenvalues with the second largest real part, that determine the relaxation time and the dynamical exponent $z$, belong to these new levels. Actually, these eigenvalues are given by a complex conjugated pair by choosing $m=1$, $m'=0$ and $P=\frac{2\pi}{L}=\frac{2\pi}{5}$ and by choosing $m=L-n_2-1=3$, $m'=1$ and $P=(L-1)\frac{2\pi}{L}=4\frac{2\pi}{5}$.

\begin{table}[ht]
\caption{The eigenvalues for $L=5$, $n_1=2$, $n_2=1$ and $n_3=1$}\label{table3}
\vspace{3mm}
\centering
\begin{scriptsize}
\begin{tabular}{|c|c|c|}
\hline
Sector   & Energies & Bethe roots  \\ 
\hline
\multirow{2}{*}{$P=0$, $m=0$, $m'=0$} & $0$ & $1$; $1$  \\ 
   & $-2$ & $i$; $-i$ \\  
\hline
\multirow{2}{*}{$P=\frac{2\pi}{5}$, $m=0$, $m'=0$} & $-0.7102901+0.3268955i$ & $0.7044758+0.2260418i$; $0.0049626-1.3516126i$  \\ 
   & $-3.0987269-0.9146808i$ & $-0.1522713+0.9227363i$; $-1.0571671-0.1604367i$  \\  
\hline
\multirow{2}{*}{$P=2\frac{2\pi}{5}$, $m=0$, $m'=0$} & $-1.2113723+1.4418449i$ & $-0.4434074-1.1954323i$; $0.6528893-0.4345885i$  \\ 
   & $-1.4796107-0.4907884i$ & $0.6498509+0.3220402i$; $-1.3593328-0.2308612i$  \\  
\hline
\multirow{2}{*}{$P=3\frac{2\pi}{5}$, $m=0$, $m'=0$} & $-1.2113723-1.4418449i$ & $-0.4434074+1.1954323i$; $0.6528893+0.4345885i$  \\ 
   & $-1.4796107+0.4907884i$ & $0.6498509-0.3220402i$; $-1.3593328+0.2308612i$  \\  
\hline
\multirow{2}{*}{$P=4\frac{2\pi}{5}$, $m=0$, $m'=0$} & $-0.7102901-0.3268955i$ & $0.7044758-0.2260418i$; $0.0049626+1.3516126i$  \\ 
   & $-3.0987269+0.9146808i$ & $-0.1522713-0.9227363i$; $-1.0571671+0.1604367i$  \\  
\hline
\multirow{2}{*}{$P=0$, $m=0$, $m'=1$} & $-0.9146286+0.4740955i$ & $0.6805844+0.0435103i$; $-0.8126913-1.2205172i$  \\ 
   & $-2.5853714-1.3401209i$ & $0.2122293+0.8634034i$; $-1.0801224+0.3136036i$ \\  
\hline
\multirow{2}{*}{$P=\frac{2\pi}{5}$, $m=0$, $m'=1$} & $-0.1515123+0.6102038i$ & $0.7626883-0.0415504i$; $0.9276613-0.9238376i$  \\ 
   & $-1.9349423-0.2034672i$ & $0.3562835+0.8258711i$; $-0.4639570-1.0103645i$  \\  
\hline
\multirow{2}{*}{$P=2\frac{2\pi}{5}$, $m=0$, $m'=1$} & $-0.5999763-0.1164667i$ & $0.5948634+0.9609114i$; $0.7314932-0.4978679i$  \\ 
   & $-3.3781712+0.3243784i$ & $-0.8574914+0.5560440i$; $-0.5334672-0.8202624i$  \\  
\hline
\multirow{2}{*}{$P=3\frac{2\pi}{5}$, $m=0$, $m'=1$} & $-0.5812538-1.1175571i$ & $0.7168460+0.2139355i$; $0.2462898+1.3138552i$  \\ 
   & $-1.7496157+0.3744123i$ & $-0.9667610+0.7800852i$; $0.5682275-0.5702092i$  \\  
\hline
\multirow{2}{*}{$P=4\frac{2\pi}{5}$, $m=0$, $m'=1$} & $-1.1838852-0.5304177i$ & $0.6689977+0.1247276i$; $-1.3569984+0.5637791i$  \\ 
   & $-1.9206433+1.5249396i$ & $-0.8877912-0.7977974i$; $0.4931159-0.6773195i$  \\  
\hline
\multirow{2}{*}{$P=0$, $m=0$, $m'=2$} & $-0.9146286-0.4740955i$ & $0.6805844-0.0435103i$; $-0.8126913+1.2205172i$  \\ 
   & $-2.5853714+1.3401209i$ & $-1.0801224-0.3136036i$; $0.2122293-0.8634033i$ \\  
\hline
\multirow{2}{*}{$P=\frac{2\pi}{5}$, $m=0$, $m'=2$} & $-1.1838852+0.5304177i$ & $0.6689977-0.1247276i$; $-1.3569984-0.5637791i$  \\ 
   & $-1.9206432-1.5249396i$ & $0.4931159+0.6773194i$; $-0.8877913+0.7977974i$  \\  
\hline
\multirow{2}{*}{$P=2\frac{2\pi}{5}$, $m=0$, $m'=2$} & $-0.5812538+1.1175571i$ & $0.7168460-0.2139355i$; $0.2462898-1.3138552i$  \\ 
   & $-1.7496155-0.3744122i$ & $0.5682275+0.5702092i$; $-0.9667610-0.7800853i$  \\  
\hline
\multirow{2}{*}{$P=3\frac{2\pi}{5}$, $m=0$, $m'=2$} & $-0.5999764+0.1164667i$ & $0.7314931+0.4978680i$; $0.5948633-0.9609114i$  \\ 
   & $-3.3781713-0.3243784i$ & $-0.5334673+0.8202623i$; $-0.8574914-0.5560439i$  \\  
\hline
\multirow{2}{*}{$P=4\frac{2\pi}{5}$, $m=0$, $m'=2$} & $-0.1515123-0.6102038i$ & $0.7626883+0.0415504i$; $0.9276613+0.9238376i$  \\ 
   & $-1.9349423+0.2034672i$ & $-0.4639570+1.0103645i$; $0.3562835-0.8258711i$  \\  
\hline
\multirow{2}{*}{$P=0$, $m=1$, $m'=0$} & $-0.6730048+0.2787791i$ & $0.7126887+0.2811734i$; $0.1922387-1.2909955i$  \\ 
   & $-3.1930207-0.7787791i$ & $-0.2488952+0.9128580i$; $-1.0220576-0.2690613i$ \\  
\hline
\multirow{2}{*}{$P=\frac{2\pi}{5}$, $m=1$, $m'=0$} & $-0.0095876-0.1567928i$ & $0.8542110-0.0537788i$; $1.1253051+0.3142423i$  \\ 
   & $-1.9958904+0.0522644i$ & $-0.1080056+1.0208495i$; $0.1011590-0.9688728i$  \\  
\hline
\multirow{2}{*}{$P=2\frac{2\pi}{5}$, $m=1$, $m'=0$} & $-0.7535156-0.3709970i$ & $0.6970108-0.1760220i$; $-0.1977531+1.3768984i$  \\ 
   & $-2.9896292+1.0401277i$ & $-1.0811538+0.0468434i$; $-0.0567203-0.9223284i$  \\  
\hline
\multirow{2}{*}{$P=3\frac{2\pi}{5}$, $m=1$, $m'=0$} & $-1.3863511-1.4870626i$ & $0.6242368+0.4958318i$; $-0.5799899+1.1122610i$  \\ 
   & $-1.4057371+0.5089149i$ & $0.6576010-0.2680371i$; $-1.4074815+0.0448287i$  \\  
\hline
\multirow{2}{*}{$P=4\frac{2\pi}{5}$, $m=1$, $m'=0$} & $-1.0413131+1.3814181i$ & $0.6752325-0.3750635i$; $-0.2901255-1.2617289i$  \\ 
   & $-1.5519502-0.4678727i$ & $0.6383338+0.3796151i$; $-1.2858348-0.3995129i$  \\  
\hline
\multirow{2}{*}{$P=0$, $m=1$, $m'=1$} & $-1.2570659-0.5290855i$ & $0.6663053+0.1698506i$; $-1.4092394+0.3592349i$  \\ 
   & $-1.7429341+1.5290855i$ & $-0.8018074-0.9104995i$; $0.5447415-0.6185860i$ \\  
\hline
\multirow{2}{*}{$P=\frac{2\pi}{5}$, $m=1$, $m'=1$} & $-0.8561485+0.4450459i$ & $0.6850761+0.0857982i$; $-0.6152818-1.3111923i$  \\ 
   & $-2.7316367-1.2540628i$ & $0.1266103+0.8921177i$; $-1.0932067+0.1912369i$  \\  
\hline
\multirow{2}{*}{$P=2\frac{2\pi}{5}$, $m=1$, $m'=1$} & $-0.0857230+0.4635491i$ & $0.7818977-0.0047923i$; $1.0392523-0.7453724i$  \\ 
   & $-1.9632204-0.1545321i$ & $0.2799188+0.8805384i$; $-0.3409951-1.0271761i$  \\  
\hline
\multirow{2}{*}{$P=3\frac{2\pi}{5}$, $m=1$, $m'=1$} & $-0.6175870-0.1729136i$ & $0.4918745+1.0873199i$; $0.7281574-0.4146480i$  \\ 
   & $-3.3334697+0.4819306i$ & $-0.9214012+0.4677083i$; $-0.4406702-0.8616120i$  \\  
\hline
\multirow{2}{*}{$P=4\frac{2\pi}{5}$, $m=1$, $m'=1$} & $-0.7242529-1.2183559i$ & $0.7058518+0.2643779i$; $0.0586461+1.3254225i$  \\ 
   & $-1.6879620+0.4093389i$ & $-1.0844991+0.6747096i$; $0.5987696-0.5044365i$  \\  
\hline
\multirow{2}{*}{$P=0$, $m=1$, $m'=2$} & $-0.2349507-0.7502975i$ & $0.7483302+0.0808009i$; $0.7839712+1.0726279i$  \\ 
   & $-1.8990239+0.2502975i$ & $-0.5896583+0.9786229i$; $0.5447415-0.6185860i$ \\  
\hline
\multirow{2}{*}{$P=\frac{2\pi}{5}$, $m=1$, $m'=2$} & $-0.9772050-0.4973541i$ & $0.6769329-0.0020008i$; $-0.9917103+1.0948804i$  \\ 
   & $-2.4295315+1.4108995i$ & $-1.0535823-0.4366956i$; $0.2924923-0.8265846i$  \\  
\hline
\multirow{2}{*}{$P=2\frac{2\pi}{5}$, $m=1$, $m'=2$} & $-1.1124249+0.5256078i$ & $0.6714067-0.0814666i$; $-1.2684629-0.7597094i$  \\ 
   & $-2.0954867-1.5037553i$ & $0.4335917+0.7324188i$; $-0.9579927+0.6801679i$  \\  
\hline
\multirow{2}{*}{$P=3\frac{2\pi}{5}$, $m=1$, $m'=2$} & $-0.4510354+1.0052159i$ & $0.7267522-0.1667177i$; $0.4348678-1.2686880i$  \\ 
   & $-1.8058199-0.3360853i$ & $0.5291823+0.6366906i$; $-0.8431292-0.8649373i$  \\  
\hline
\multirow{2}{*}{$P=4\frac{2\pi}{5}$, $m=1$, $m'=2$} & $-0.5893425+0.0585923i$ & $0.7266229+0.5951587i$; $0.6653905-0.8311391i$  \\ 
   & $-3.4051795-0.1631208i$ & $-0.6226556+0.7686536i$; $-0.7857321-0.6360565i$  \\  
\hline
\end{tabular}
\end{scriptsize}
\end{table}

\begin{table}[ht]
\caption{The eigenvalues for $L=5$, $n_1=2$, $n_2=1$ and $n_3=1$}\label{table3b}
\vspace{3mm}
\centering
\begin{scriptsize}
\begin{tabular}{|c|c|c|}
\hline
Sector   & Energies & Bethe roots  \\ 
\hline
\multirow{2}{*}{$P=0$, $m=2$, $m'=0$} & $-0.8782556+1.3066224i$ & $0.6924404-0.3181480i$; $-0.1217444-1.3066224i$  \\ 
   & $-1.6217444-0.4405970i$ & $0.6217444+0.4405970i$; $-1.1924404-0.5478775i$ \\  
\hline
\multirow{2}{*}{$P=\frac{2\pi}{5}$, $m=2$, $m'=0$} & $-0.6420067+0.2272404i$ & $0.7210291+0.3432755i$; $0.3565184-1.2004077i$  \\ 
   & $-3.2715388-0.6339770i$ & $-0.3454212+0.8925020i$; $-0.9765410-0.3717780i$  \\  
\hline
\multirow{2}{*}{$P=2\frac{2\pi}{5}$, $m=2$, $m'=0$} & $-0.0382564-0.3118707i$ & $0.8094973-0.0284654i$; $1.1094939+0.5414704i$  \\ 
   & $-1.9835959+0.1039590i$ & $-0.2219898+1.0302847i$; $0.1946917-0.9286414i$  \\  
\hline
\multirow{2}{*}{$P=3\frac{2\pi}{5}$, $m=2$, $m'=0$} & $-0.8022829-0.4105415i$ & $0.6905308-0.1296573i$; $-0.4074337+1.3637265i$  \\ 
   & $-2.8668478+1.1536863i$ & $-1.0933657-0.0707271i$; $0.0366096-0.9119650i$  \\  
\hline
\multirow{2}{*}{$P=4\frac{2\pi}{5}$, $m=2$, $m'=0$} & $-1.3312716+0.5218134i$ & $0.6627602-0.2174407i$; $-1.4253665-0.1539334i$  \\ 
   & $-1.5641999-1.5163354i$ & $0.5883753+0.5576163i$; $-0.6993882+1.0161915i$  \\  
\hline
\multirow{2}{*}{$P=0$, $m=2$, $m'=1$} & $-0.8782556-1.3066224i$ & $0.6924404+0.3181480i$; $-0.1217444+1.3066224i$  \\ 
   & $-1.6217444+0.4405970i$ & $0.6217444-0.4405970i$; $-1.1924404+0.5478775i$ \\  
\hline
\multirow{2}{*}{$P=\frac{2\pi}{5}$, $m=2$, $m'=1$} & $-1.3312716-0.5218134i$ & $0.6627602+0.2174407i$; $-1.4253665+0.1539334i$  \\ 
   & $-1.5641999+1.5163354i$ & $-0.6993882-1.0161915i$; $0.5883753-0.5576162i$  \\  
\hline
\multirow{2}{*}{$P=2\frac{2\pi}{5}$, $m=2$, $m'=1$} & $-0.8022829+0.4105415i$ & $0.6905308+0.1296573i$; $-0.4074337-1.3637265i$  \\ 
   & $-2.8668476-1.1536863i$ & $0.0366097+0.9119649i$; $-1.0933658+0.0707270i$  \\  
\hline
\multirow{2}{*}{$P=3\frac{2\pi}{5}$, $m=2$, $m'=1$} & $-0.0382564+0.3118707i$ & $0.8094973+0.0284654i$; $1.1094939-0.5414704i$  \\ 
   & $-1.9835961-0.1039590i$ & $0.1946916+0.9286414i$; $-0.2219899-1.0302847i$  \\  
\hline
\multirow{2}{*}{$P=4\frac{2\pi}{5}$, $m=2$, $m'=1$} & $-0.642007-0.2272403i$ & $0.3565184+1.2004077i$; $0.7210291-0.3432755i$  \\ 
   & $-3.2715388+0.6339770i$ & $-0.9765410+0.3717780i$; $-0.3454213-0.8925020i$  \\  
\hline
\multirow{2}{*}{$P=0$, $m=2$, $m'=2$} & $-0.5857864$ & $0.7071068+0.7071068i$; $0.7071068-0.7071068i$  \\ 
   & $-3.4142136$ & $-0.7071068+0.7071068i$; $-0.7071068-0.7071068i$ \\  
\hline
\multirow{2}{*}{$P=\frac{2\pi}{5}$, $m=2$, $m'=2$} & $-0.3351580-0.8824028i$ & $0.7368286+0.1224476i$; $0.6168508+1.1882338i$  \\ 
   & $-1.8558250+0.2946176i$ & $-0.7166508+0.9306243i$; $0.4810054-0.7024638i$  \\  
\hline
\multirow{2}{*}{$P=2\frac{2\pi}{5}$, $m=2$, $m'=2$} & $-1.0433294-0.5145741i$ & $0.6739498+0.0394376i$; $-1.1454536+0.9391785i$  \\ 
   & $-2.2656875+1.4656306i$ & $-1.0130569-0.5593159i$; $0.3665267-0.7825715i$  \\  
\hline
\multirow{2}{*}{$P=3\frac{2\pi}{5}$, $m=2$, $m'=2$} & $-1.0433294+0.5145741i$ & $0.6739498-0.0394376i$; $-1.1454536-0.9391785i$  \\ 
   & $-2.2656875-1.4656306i$ & $-1.0130569+0.5593159i$; $0.3665267+0.7825715i$  \\  
\hline
\multirow{2}{*}{$P=4\frac{2\pi}{5}$, $m=2$, $m'=2$} & $-0.3351580+0.8824028i$ & $0.7368286-0.1224476i$; $0.6168508-1.1882338i$  \\ 
   & $-1.8558250-0.2946176i$ & $0.4810054+0.7024638i$; $-0.7166509-0.9306242i$  \\  
\hline
\multirow{2}{*}{$P=0$, $m=3$, $m'=0$} & $-1.2570659+0.5290855i$ & $0.6663053-0.1698506i$; $-1.4092394-0.3592349i$  \\ 
   & $-1.7429341-1.5290855i$ & $0.5447415+0.6185860i$; $-0.8018074+0.9104995i$ \\  
\hline
\multirow{2}{*}{$P=\frac{2\pi}{5}$, $m=3$, $m'=0$} & $-0.7242529+1.2183560i$ & $0.7058518-0.2643779i$; $0.0586461-1.3254224i$  \\ 
   & $-1.6879619-0.4093389i$ & $0.5987696+0.5044365i$; $-1.0844991-0.6747097i$  \\  
\hline
\multirow{2}{*}{$P=2\frac{2\pi}{5}$, $m=3$, $m'=0$} & $-0.6175868+0.1729136i$ & $0.7281573+0.4146480i$; $0.4918745-1.0873199i$  \\ 
   & $-3.3334696-0.4819305i$ & $-0.4406701+0.8616120i$; $-0.9214011-0.4677083i$  \\  
\hline
\multirow{2}{*}{$P=3\frac{2\pi}{5}$, $m=3$, $m'=0$} & $-0.0857230-0.4635491i$ & $0.7818977+0.0047922i$; $1.0392523+0.7453723i$  \\ 
   & $-1.9632206+0.1545321i$ & $-0.3409952+1.0271760i$; $0.2799187-0.8805385i$  \\  
\hline
\multirow{2}{*}{$P=4\frac{2\pi}{5}$, $m=3$, $m'=0$} & $-0.8561485-0.4450459i$ & $0.6850761-0.0857982i$; $-0.6152818+1.3111923i$  \\ 
   & $-2.7316368+1.2540628i$ & $-1.0932067-0.1912369i$; $0.1266102-0.8921177i$  \\  
\hline
\multirow{2}{*}{$P=0$, $m=3$, $m'=1$} & $-0.6730048-0.2787791i$ & $0.7126887-0.2811734i$; $0.1922387+1.2909955i$  \\ 
   & $-3.1930206+0.7787790i$ & $-1.0220576+0.2690613i$; $-0.2488952-0.9128580i$ \\  
\hline
\multirow{2}{*}{$P=\frac{2\pi}{5}$, $m=3$, $m'=1$} & $1.0413131-1.3814181i$ & $0.6752325+0.3750635i$; $-0.2901255+1.2617289i$  \\ 
   & $-1.5519502+0.4678727i$ & $-1.2858348+0.3995130i$; $0.6383338-0.3796150i$  \\  
\hline
\multirow{2}{*}{$P=2\frac{2\pi}{5}$, $m=3$, $m'=1$} & $-1.4057371-0.5089149i$ & $0.6576010+0.2680371i$; $-1.4074815-0.0448287i$  \\ 
   & $-1.3863511+1.4870625i$ & $-0.5799900-1.1122609i$; $0.6242368-0.4958318i$  \\  
\hline
\multirow{2}{*}{$P=3\frac{2\pi}{5}$, $m=3$, $m'=1$} & $-0.7535156+0.3709970i$ & $0.6970108+0.1760220i$; $-0.1977531-1.3768984i$  \\ 
   & $-2.9896293-1.0401277i$ & $-0.0567204+0.9223285i$; $-1.0811538-0.0468433i$  \\  
\hline
\multirow{2}{*}{$P=4\frac{2\pi}{5}$, $m=3$, $m'=1$} & $-0.0095876+0.1567928i$ & $0.8542110+0.0537788i$; $1.1253051-0.3142423i$  \\ 
   & $-1.9958905-0.0522644i$ & $0.1011589+0.9688728i$; $-0.1080056-1.0208495i$  \\  
\hline
\multirow{2}{*}{$P=0$, $m=3$, $m'=2$} & $-0.2349507+0.7502975i$ & $0.7483302-0.0808009i$; $0.7839712-1.0726279i$  \\ 
   & $-1.8990239-0.2502976i$ & $0.4233823+0.7660263i$; $-0.5896583-0.9786229i$ \\  
\hline
\multirow{2}{*}{$P=\frac{2\pi}{5}$, $m=3$, $m'=2$} & $-0.5893423-0.0585923i$ & $0.6653905+0.8311391i$; $0.7266229-0.5951587i$  \\ 
   & $-3.4051794+0.1631208i$ & $-0.7857321+0.6360565i$; $-0.6226555-0.7686536i$  \\  
\hline
\multirow{2}{*}{$P=2\frac{2\pi}{5}$, $m=3$, $m'=2$} & $-0.4510354-1.0052159i$ & $0.7267522+0.1667177i$; $0.4348678+1.2686880i$  \\ 
   & $-1.8058198+0.3360852i$ & $0.5291824-0.6366905i$; $-0.8431291+0.8649374i$  \\  
\hline
\multirow{2}{*}{$P=3\frac{2\pi}{5}$, $m=3$, $m'=2$} & $-1.1124249-0.5256078i$ & $0.6714067+0.0814665i$; $-1.2684629+0.7597094i$  \\ 
   & $-2.0954868+1.5037554i$ & $-0.9579926-0.6801680i$; $0.4335916-0.7324189i$  \\  
\hline
\multirow{2}{*}{$P=4\frac{2\pi}{5}$, $m=3$, $m'=2$} & $-0.9772050+0.4973541i$ & $0.6769329+0.0020008i$; $-0.9917102-1.0948804i$  \\ 
   & $-2.4295316-1.4108996i$ & $0.2924923+0.8265847i$; $-1.0535823+0.4366957i$  \\  
\hline
\end{tabular}
\end{scriptsize}
\end{table}

\end{document}